\documentclass[a4paper,12pt]{article}
\pdfoutput=1
\usepackage{amsmath,amsfonts,amssymb,epsfig}
\usepackage{slashed}
\usepackage{cite}
\usepackage[width=0.9\textwidth,margin={30pt,30pt},font=small,labelfont=bf]{caption}
\numberwithin{equation}{section}

\usepackage{hyperref}
\usepackage{subfig}

\usepackage{xcolor}
\hypersetup{
    colorlinks,
    linkcolor={red!0!black},
    citecolor={blue!50!black},
    urlcolor={blue!80!black}
}

\usepackage{graphicx}
\usepackage{cite}
\usepackage{color}
\usepackage{bm}
\usepackage{multirow}

\usepackage{xspace}

\newcommand{\scr}[1]{\ensuremath{\mathcal{#1}}}

\def\nn{\nonumber}

\def\d{\partial}

\def\vep{\varepsilon}
\def\s{\sigma}

\def\L{\scr{L}}
\def\eb{\mathbf{e}}
\def\atw{\alpha_{\! \!{}_\text{ tw}} }
\def\phifv{\phi_{\text{\tiny F}} }
\def\phitv{\phi_{\text{\tiny T}} }
\def\phiout{ \varphi_{ \text{\tiny out}}}
\def\phiin{ \varphi_{ \text{\tiny in}}}
\def\out{ \text{\tiny out}}
\def\in{ \text{\tiny in}}

\def\nn{\nonumber}

\newcommand\blfootnote[1]{%
  \begingroup
  \renewcommand\thefootnote{}\footnote{#1}%
  \addtocounter{footnote}{-1}%
  \endgroup
}

\begin{document}

\title{\Large  {\bf Towards the fate of the oscillating false vacuum}}

\author{\\[2em]  Luc Darm\'e,$^{1,a}$\blfootnote{$^a$  \url{luc.darme@ncbj.gov.pl}} Joerg Jaeckel$^{2,b}$\blfootnote{ $^b$ \url{jjaeckel@thphys.uni-heidelberg.de}}  \let\theblfootnote\relax and  Marek Lewicki,$^{3,4,c}$\blfootnote{$^c$ \url{marek.lewicki@fuw.edu.pl}}\\[5ex]
\small {\em $^1$ National Centre for Nuclear Research,}\\
\small {\em Ho{\. z}a 69, 00-681 Warsaw, Poland }\\
\small {\em $^2$ Institut f\"ur Theoretische Physik, Universit\"at Heidelberg,}\\
\small {\em Philosophenweg 16, 69120 Heidelberg, Germany,}\\
\small {\em $^3$ ARC Centre of Excellence for Particle Physics at the Terascale (CoEPP) \& CSSM,  }\\
\small {\em Department of Physics, University of Adelaide, South Australia 5005, Australia}\\
\small {\em $^4$ Faculty of Physics, University of Warsaw ul.\ Pasteura 5, 02-093 Warsaw, Poland}
}
%
\date{}
\maketitle

\abstract{Motivated by cosmological examples we study quantum field theoretical tunnelling from an initial state where the ``classical field'', i.e. the vacuum expectation value of the field operator is spatially homogeneous but performing a time-dependent oscillation about a local minimum. In particular we estimate both analytically and numerically the exponential contribution to the tunnelling probability. We additionally show that after the tunnelling event, the classical field solution -- the so-called ``bubble'' -- mediating the phase transition can either grow or collapse. We present a simple analytical criterium to distinguish between the two behaviours.}
\newpage 

\tableofcontents

\setcounter{footnote}{0}

\section{Introduction}

The decay probability of a scalar field settled in a local minimum of its potential to its global minimum is a fascinating process which appeals more to the quantum aspects of Quantum Field Theory than its classical ones. 

In a seminal paper of Coleman~\cite{Coleman:1977py} it was understood that the tunnelling rate could be obtained by studying the so-called ``Most Probable Escape Path'' (MPEP) introduced in~\cite{banks_coupled_1973,banks_coupled_1973-1}. This path corresponds to a solution of the quantum equations of motion in Euclidean time and gives the trajectory in field space along which the tunnelling probability is maximal. Subsequent calculations of the first-order quantum corrections to this path by Coleman and Callan in~\cite{Callan:1977pt} led to the famous formula for the tunnelling rate: 
\begin{equation}
 \frac{\Gamma}{V} = A e^{-S} (1+ \scr{O}(\hbar)) \ .
 \label{eq:CLdecayrate}
\end{equation}
Here, $S$ is twice the imaginary part of the action required to form a bubble, i.e. the smallest action configuration connecting the two vacua and therefore the most probable escape path. $A$ is the dimensionful prefactor that can be determined by a loop calculation around the bubble solution and was first estimated in~\cite{Callan:1977pt}.

This result has been since extended to include thermal~\cite{Linde:1981zj,Garriga:1994ut,Ferrera:1995gs} and gravitational~\cite{Coleman:1980aw} corrections (see~\cite{Espinosa:2015qea,Rose:2015lna,Salvio:2016mvj,Rajantie:2016hkj,Czerwinska:2016fky} for a modern use of these techniques in the Higgs boson case). Furthermore, much interest in phase transitions which could occur in the early universe was sparked recently by the potentially detectable gravitational wave signals they could leave behind~\cite{Beniwal:2017eik,Artymowski:2016tme,Kobakhidze:2016mch,Kobakhidze:2017mru,Marzola:2017jzl,Vaskonen:2016yiu,Kakizaki:2015wua,Hashino:2016xoj,Balazs:2016tbi}.
These situations have in common that they start from a classically stable and essentially time-independent situation.
The only time dependence arises from the tunnelling process itself. However, in many cosmological situations the Universe is not in a quasi-equilibrium and the field relevant to the tunnelling process may be time-dependent.   
A recent example of such a situation arises at the end of monodromy inflation~\cite{Silverstein:2008sg,McAllister:2008hb}. Here the inflaton field may roll through a potential featuring several local minima with the Hubble expansion slowing down the evolution of the field, eventually trapping it in one of the minima~\cite{Hebecker:2016vbl}. 
Yet, even after being classically trapped in one minimum the field continues to oscillate. Naturally the question arises what is the probability for a transition from this oscillating state to one of the neighbouring minima, in particular since this has interesting phenomenological consequences in the form of gravitational waves~\cite{Hebecker:2016vbl}.
Another example are models of cosmological relaxation~\cite{Graham:2015cka,Espinosa:2015eda,Hardy:2015laa,Patil:2015oxa,Jaeckel:2015txa,Gupta:2015uea,Batell:2015fma} and models of dark matter based on axion-monodromies~\cite{Jaeckel:2016qjp} where again we have evolution in a potential with many minima and the field continuing to oscillate for a significant time after it has been trapped in one of them.

For these reasons we are interested in obtaining a better understanding of tunnelling from a time-dependent initial state. As a simple first step we will consider in the present paper tunnelling of a scalar field \textit{classically} oscillating around a false vacuum, as motivated by the above examples.
Our aim is to provide first estimates for the tunnelling rate from such a non-trivial state. This has already been partially studied in~\cite{KeskiVakkuri:1996gn} and we will compare our results to theirs and comment on the differences.
Calculations have been performed also for a variety of other non-trivial initial states, for instance see~\cite{Hiscock:1987hn,Gregory:2013hja,Burda:2015yfa,Burda:2016mou} in the case of black holes and~e.g.\cite{Monin:2010zz} discussing tunnelling from particle collisions.

Let us start with some qualitative expectations. As the energy in the oscillating state is higher than in the situation when the field sits at its minimum it is natural to expect that the tunnelling rate will be increased, and will have additionally a time dependence. Furthermore, since the oscillating state breaks the $SO(3,1)$ boost symmetries, one should expect the emergence of other types of ``bubble'' solutions besides the usual expanding Coleman-De Luccia instanton. Finally, even if the minimum around which the field oscillates is the global minimum, it seems possible that for sufficiently large oscillations the field actually transits into a neighbouring potential well whose minimum is higher as long as the oscillations are big enough (in quantum mechanics transitions under all these circumstances are possible and do occur).

In the present paper we calculate the exponential term $S$ of the decay rate~\eqref{eq:CLdecayrate} and investigate whether or not the bubble arising from the MPEP will subsequently generate a phase transition, or instead collapse after its nucleation. Interestingly, and in accordance with the naive expectation, we find that on average, the tunnelling rate is dominated by its value at an extremum of the oscillation:
\begin{align*}
\big\langle  ~\frac{\Gamma}{V} ~ \big\rangle  ~\propto ~ \exp \left[ - S_{ext} \right] \ ,
\end{align*} 
where $S_{ext}$ is twice the imaginary part of the action required to form a bubble of true vacuum when the oscillating field is at an extremum of its oscillation. However, for oscillations large enough, the bubble created by the MPEP at the turning point of the oscillation will collapse, so that the true decay rate should be suppressed compared to the above expectation. This happens when:
\begin{align*}
 \varphi_{\out}^2 > \frac{\Delta V }{\mu^2} \ ,
\end{align*}
where, $ \varphi_{\out}$ is the amplitude of the initial oscillations, $\Delta V$ is the energy difference between the two minima and $\mu$ is the so-called ``inverse thickness'' of the wall, defined in Sect.~\ref{thinwall}. This equation also suggests that while is possible to tunnel from the global minimum to a local one, the bubble a false vacuum will always collapse, falling short of generating a phase transition. 

These results, along with their corresponding caveats and a more precise numerical study, will be presented in the rest of this paper. Let us nevertheless simply comment on how a phase transition can occur when the oscillations are larger than the previously mentioned criterium. It is clear that a phase transition can only occur in this case from bubbles which do not originate from the MPEP. For instance, the field inside the bubble could have either a non-vanishing time derivative when the bubble nucleates, or simply a value different from the true vacuum one, triggering subsequent oscillations within the bubble. We will investigate these possibilities both numerically and analytically in this work. Notice that one cannot rule out that more exotic configurations could provide the dominant contribution in this case.

While the exponential term of the average decay rate depends only of the tunnelling rate from an extremum of the oscillation, calculating the precise time-dependence of the tunnelling rate is a more delicate matter. We present in this work an intuitive estimation of this dependence. We leave to a future work~\cite{2Dreduction} a more precise and theoretically sound study of this effect based on the  Functional Schr\"odinger Equation method.

In the following we will proceed as follows. In Sect.~\ref{thinwall} we will setup the explicit model we study and briefly recall the essentials of tunnelling and in particular the thin wall approximation. The latter being very useful for obtaining analytical estimates. We will then modify the thin-wall approximation to suit our non-trivial initial state and pursue both estimates described above. We then compare the thin-wall results to a numerical calculation of bubbles and their time evolution in Sect.~\ref{numerical}. Finally in Sect.~\ref{conclusions} we summarise our findings and draw some conclusions.

\section{False vacuum decay and membrane action}\label{thinwall}

Most analytical results dealing with tunnelling in QFT are based on the so-called ``thin-wall limit''. We start by reviewing in this section the standard Coleman approach to tunnelling in QFT and introduce this thin-wall limit. A particularly interesting aspect of the thin-wall regime is the possibility of simplifying the problem to the nucleation and subsequent evolution of a ``bubble'' of true vacuum. This is described by a ``membrane'' action that we will also present in more details below.

\subsection{The fate of the false vacuum}

\begin{figure}
\begin{center}
     \includegraphics[width=0.5\textwidth]{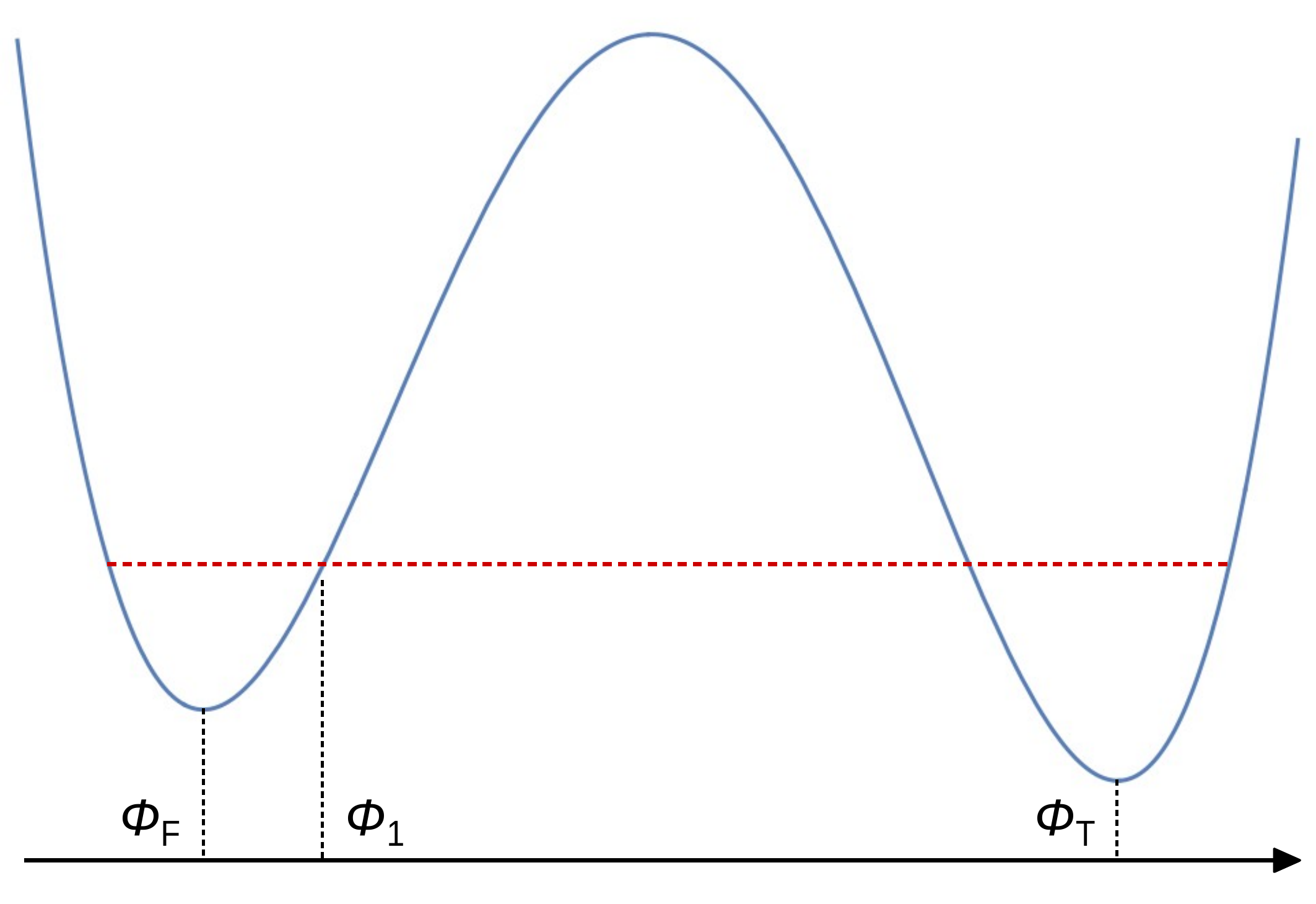}
 \caption{Schematic form of the potential considered in this work.}
 \label{fig:oscillating}
\end{center}
\end{figure}

For concreteness, most of the results in this paper will be derived using an asymmetric double well potential of the form 
\begin{align}
 \label{eq:potential}
 V = \frac{g c^4}{4} (\phi^2/c^2 - 1)^2 - b (\phi + c) \ ,
\end{align}
where $g,c$ and $b$ are positive constants. This potential has two minima in $\phifv$ and $\phitv$, with the latter being deeper with an asymmetry of approximately $ 2 b c$. Note that more generally, most of the qualitative properties we will derive will be also true for all potentials of the general form of Figure~\ref{fig:oscillating}.
Anticipating slightly, we will furthermore define the inverse ``thickness'' of the wall
\begin{align}
 \mu \equiv \sqrt{2 g c^2} \ ,
\end{align}
and assume the following hierarchy, called thin-wall limit in the rest of this paper:
\begin{align}
\label{eq:thinwall}
 b\, c  \ll \mu^2 c^2 \ .
\end{align}
It is finally useful to introduce a small parameter for this hierarchy~\eqref{eq:thinwall}, $\atw$ defined by:
\begin{align}\label{eq:alpha}
 \atw \equiv \frac{b}{\mu^2 c} \ .
\end{align}

In the approach of Coleman to vacuum-to-vacuum tunnelling, the exponential contribution of~\eqref{eq:CLdecayrate} can be obtained by calculating the MPEP: a path in field space corresponding to a saddle point of the Euclidean action
\begin{align}\label{eq:action0}
 S_E = \int d \tau d^3 x \left[ \frac{1}{2} (\d_\tau \phi) ^2 +  \frac{1}{2} (\nabla \phi) ^2 + V(\phi)\right] \ ,
\end{align}
which matches the false vacuum at spatial infinity limit, $\phi \rightarrow \phifv $ at $|\vec{x}| \rightarrow \infty $. Supposing an $O(4)$-symmetric solution, we define $r= \sqrt{\tau^2+|\vec{x}|^2}$ and write the Euclidean equations of motion as:
\begin{align}\label{eq:EOM0}
 \frac{\d^2 \phi}{\d r^2} +  \frac{3}{r} \frac{\d \phi}{\d r} = V'(\phi) \ .
\end{align}
Solving this equation leads to the ``Coleman bounce'' solution. This can be matched with the classical solitonic solution after the tunnelling event is done at Euclidean time $\tau =0$. Here $\d_\tau \phi = 0$ so that one can smoothly rotate the Euclidean time into real time, at which point the field evolves classically again. This simplified story has been given a sound theoretical basis, both in terms of a path integral~\cite{Callan:1977pt,Coleman:1980aw}, and in terms of a Functional Schr\"odinger Equation~\cite{Gervais:1977nv,Bitar:1978vx} approach.

In the thin wall limit~\eqref{eq:thinwall}, one can neglect the viscous damping first derivative term, leading for the potential~\eqref{eq:potential} to the usual bounce solution
\begin{align}
\label{eq:phibounce} 
 \phi_0(r) = -c \tanh \left(\frac{\mu}{2} (r-R_0) \right) \ ,
\end{align}
where $R_0$ is a constant which will be fixed by energy conservation. The fact that $R_0$ is not imposed by the Euclidian equation of motion is a consequence of our choice of neglecting the first derivative term. Indeed, complete calculations for simplified, piecewise potentials (see, e.g~\cite{Duncan:1992ai,Theodorakis:1999vr,deSouzaDutra:2009sc,Correa:2015rka}) fix the initial radius.

\subsection{Membrane action}

In the thin-wall limit, we can neglect the details of the potential shape and parametrise the whole evolution of the system as a function of the membrane tension $\s$, the differential pressure $p$ between the inside and outside of the membrane and the bubble radius $R$ (we assume a spherically symmetric bubble). We will suppose that both the field inside and outside of the bubble are spatially constant (we will comment on this approximation below). 
In the thin wall limit we can now derive the membrane Lagrangian directly from the action of the scalar field (cf. App.~\ref{app:mem}), it can be expressed in term of the radius of the bubble $R$ as:
\begin{align}
\label{eq:Lmembrane}
\L_m = -4 \pi \s R^2 \sqrt{1-\dot{R}^2} + \frac{4}{3} \pi p R^3 + \frac{4}{3} \pi p_{\out} \Lambda^3 ,
\end{align}
where $\Lambda$ is the radius of a large control volume including the bubble. As we will see this part does not affect the evolution of the bubble radius $R$ and of the field inside but is nonetheless required for energy conservation. Writing the field inside (outside) the membrane $c+\phiin$ ($-c+\phiout$), we have for the pressure and tension,
\begin{align}
\label{eq:tension}
\sigma =  \int_{-c+\phiout}^{c+\phiin} d\phi \sqrt{2 V(\phi)} \ ,
\end{align}
and \begin{align}
\label{eq:pressure}
  p  \equiv p_{\in} - p_{\out} \equiv \left(\frac{1}{2} \dot{\varphi}_{\in}^2 - V(c+\phiin)\right) - \left(\frac{1}{2} \dot{\varphi}_{\out}^2 - V(-c+\phiout) \right) \ . 
\end{align}
In particular, for the potential~\eqref{eq:potential} introduced earlier, the tension can be written as:
\begin{align}
\label{eq:tension2}
\sigma =  \frac{2 \mu }{3} \left[ c^2 + \scr{O} \left(  \phiin^2,\phiout^2 \right) \right]\ ,
\end{align}
where we have used the ``thin-wall'' limit to approximate the field within and outside the walls as the two minima of the potential.  

The dominant corrections in the processes we will be interested in are of the form $\frac{\phiin^2}{c^2} \times \frac{1}{\atw}$. The thin-wall hierarchy implies that they are hierarchically larger than those arising from the anharmonicity of the potential as in~\eqref{eq:tension2}. In the rest of this paper we will therefore only retain the former and discard the latter.

An important comment about~\eqref{eq:Lmembrane} is that the observable which appears naturally in the membrane Lagrangian is the differential of pressure $p$, and not the differential energy density 
\begin{align}
 \vep \equiv \vep_{\in} - \vep_{\out} \equiv \left(\frac{1}{2} \dot{\varphi}_{\in}^2 + V(c+\phiin)\right) - \left(\frac{1}{2} \dot{\varphi}_{\out}^2 + V(-c+\phiout) \right) \ . 
\end{align}
While the two are opposite in the static case, we have more generally $ p = (\dot{\varphi}_{\in}^2 - \dot{\varphi}_{\out}^2) - \vep$.\footnote{The previous expressions hold as long as the fields are spatially constant.}

The equations of motion following from Eq.~\eqref{eq:Lmembrane} are,
\begin{eqnarray}
\label{Requation}
&&\ddot{R}+2\frac{1-\dot{R}^2}{R}=\frac{p}{\sigma}(1-\dot{R}^2)^{3/2},
\\
\label{phiinequation}
&&\ddot{\varphi}_{\in}+3\frac{\dot{\varphi}_{\rm in}\dot{R}}{R}=V'(c+\phiin) \ ,
\\
\label{phioutequation}
&&\ddot{\varphi}_{\out}-3\frac{R^2\dot{R}\dot{\varphi}_{\rm out}}{\Lambda^3-R^3}=V'(-c+\phiout).
\end{eqnarray}
As mentioned above $\varphi_{\rm out}$ decouples from the equation for the bubble radius and the field inside the bubble. Moreover, it is the only part whose evolution explicitly depends on the assumed cutoff radius $\Lambda$ hinting that the evolution outside the bubble is not completely accounted for in the present approximation.
As it doesn't affect the part we are interested in we will mostly ignore $\varphi_{\rm out}$. Let us remark, however, that the energy of the system is only conserved if one takes it into account.  Furthermore, it is clear that we have two different time scales, one of order $\mu^{-1}$  and the other of order the initial radius $R_0$. More precisely, $\varphi_{\in}$  and $R(t)$ will have both a fast oscillating behaviour of frequency $\mu$ slowly modulated over a time scale $R_0$. The hierarchy~\eqref{eq:thinwall} then implies that the former is order of magnitude smaller than the latter. This will be used in the next section to average over several field oscillations while keeping the bubble radius around $R_0$.

The Lagrangian~\eqref{eq:Lmembrane} has three degrees of freedom, $R(t),\phiin (t) $ and $\phiout (t)$ and its conserved Hamiltonian is therefore:
\begin{align}
\label{eq:Hamiltonian}
\scr{H} &= \dot{\varphi}_{\in} \frac{\d \L_m}{\d \dot{\varphi}_{\in}} +\dot{\varphi}_{\out} \frac{\d \L_m}{\d \dot{\varphi}_{\out}}+\dot{R} \frac{\d \L_m}{\d \dot{R}} - \L_m \nn \\
&=\frac{ 4 \pi \s R^2}{ \sqrt{1-\dot{R}^2}}  + \frac{4}{3} \pi \left[ (\dot{\varphi}_{\in}^2 - \dot{\varphi}_{\out}^2) - p \right] R^3  +  \frac{4}{3} \pi \vep_{\out} \Lambda^3 \nn \\
&=\frac{ 4 \pi \s R^2}{ \sqrt{1-\dot{R}^2}}  + \frac{4}{3} \pi \vep R^3 +  \frac{4}{3} \pi \vep_{\out} \Lambda^3 
\end{align}
Conservation of energy implies that this is equal to the energy of the scalar field contained in the control volume before the bubble actually nucleates, $\scr{H}_{ini} = \frac{4}{3} \pi \vep_{\out} \Lambda^3$. 
We find\footnote{After nucleation this equation neglects changes in $\epsilon_{\out}$ and is therefore only approximate.}
\begin{align}
\label{eq:Econs}
\frac{ 4 \pi \s R^2}{ \sqrt{1-\dot{R}^2}}  + \frac{4}{3} \pi \vep R^3  = 0 \ .
\end{align}
Fixing $\dot{R} = 0$ in~\eqref{eq:Econs}, and noting $\vep_0$ the differential energy density when the bubble nucleates, we see that energy conservation gives the radius $R_0$ of the classical bubble right after it nucleates:
\begin{align}
\label{eq:stab}
R_0 = \frac{3 \s}{-\vep_0} \ ,
\end{align}

In the case of a vacuum-to-vacuum tunnelling event, one can fix the value of the false vacuum to zero,  so that $\vep= \vep_0 =  V(c+\phiin)$. Writing $\tilde{R}_0$ for the initial radius in this case, Eq.~\eqref{eq:Econs} is then particularly simple and leads to
\begin{align*}
\sqrt{1-\dot{R}^2} = \frac{\tilde{R}_0}{R} \Rightarrow \dot{R} = \sqrt{1-\left( \frac{\tilde{R}_0}{R} \right)^2} \ ,
\end{align*}
where the positive sign solution is imposed by the first equality since it also implies $R>\tilde{R}_0$. We therefore recover the usual evolution of the bubble wall: 
\begin{align*}
R(t) = \sqrt{\tilde{R}_0^2 + t^2} \ .
\end{align*}

More generally~\eqref{eq:Econs} relates the evolution of the bubble radius to the differential of energy density $\vep$. An important comment here is that during the bubble evolution, $\vep$ is in fact \emph{not} constant.  This is a consequence of the spatial inhomogeneities introduced by the wall and the spatial perturbations created by its evolution. 
The explicit bubble profile can be calculated~\cite{KeskiVakkuri:1996gn}.
However, it is important to note that at early time (namely as long as we of $R \sim R_0$ up to $\atw$ corrections), the oscillating bubble profile does not significantly modify the predictions of the spatially homogeneous model we introduced above. Indeed, corrections to the tension are of subleading order $\scr{O} (\left(\frac{\phiin}{c} \right)^2),\scr{O} (\left(\frac{\phiout}{c} \right)^2 )$. Similarly, the contributions to the pressure are suppressed by factors of $\sqrt{\atw}$ compared to the dominant corrections to the vacuum-to-vacuum case.

In the analytical result of the next section, we will therefore rely on the spatially homogeneous model of~\eqref{eq:Lmembrane}. However, if one wants to describe accurately the evolution of a contracting bubble when $R < R_0$, then this simplified picture does not hold anymore because of two separate effects:
\begin{itemize}
 \item First, the contraction of the bubble wall will lead to an exponential increase of the amplitude of the oscillations within the bubble. This can be easily seen from the equation of motion~\eqref{phiinequation} for the homogeneous field inside the bubble. It will oscillate around the true vacuum with an amplification/damping term depending on the evolution of the bubble. This implies that even small initial oscillations of the field within the bubble will grow exponentially as soon as the bubble starts contracting. 
 
\item Second, while the wall is contracting, it leaves behind a region of spatial oscillations around the false vacuum. Energy conservation then implies that this region must have a lower energy density than the initial oscillating field, providing the energy necessary to contract the bubble. This is what we observed in the lattice simulations of Sect.~\ref{sec:lattice}.
 \end{itemize}

Let us close this section by commenting on the possible final fate of a contracting bubble. The thin-wall approximation clearly breaks down for very small radius where the damping term $\frac{2}{\rho} \d_\rho$ becomes non-negligible. As it was shown in~\cite{Gleiser:1993pt} (and subsequent literature), one can have in this case the appearance of solitonic quasi-periodic solutions called ``oscillons''. This depends however on the original radius of the bubble. In units of $\mu/2$, it was found that at small radius, a quasi-periodic behaviour could be found. In the case of larger contracting bubbles it was claimed that they will start with a period of quasi relativistic contraction and then settle shortly into oscillons before radiating away their energy.

\section{Decay from an oscillating state}
\label{sec:Oscillating}
In this section we will now take steps to obtain a tunnelling rate for an oscillating initial field.
We begin by applying our thin-wall approximation to the field when it is at the turning point.
In the next step we consider the evolution of the bubble. As it turns out for large enough oscillations the bubble will contract, we provide an analytical criterion distinguishing between contracting and expanding bubbles. We then turn to an estimate of the time-dependent decay rate at points in time when the field is not at a turning point. This allows us to estimate the time dependent as well as the average tunnelling rate.

\subsection{Tunnelling from the turning point}
 In the limit of small oscillations the fields are approximately harmonic oscillators of the form:
\begin{align}
\label{eq:fieldsbubble}
\phiin &= -q_{\in} c \cos\left( \mu (t - t_0)\right) \\
\phiout &= q_{\out} c \cos \left(\mu t\right) \nn \ ,
\end{align}
where the time origin has been chosen such that the field outside of the bubble is at an extremum of its oscillation at $t=0$. As in the previous section, we denote by $q_{\in}$ and $q_{\out}$ the amplitude of the oscillations, normalised to $c$.

If our tunnelling process occurs at the turning point of the oscillating field the time derivative of the field is given by $\dot{\phi}=0$ everywhere. This allows for a smooth analytical continuation to Euclidean space, making this a good starting point for our investigation.
 
Energy conservation leads to the expression for the bubble radius~\eqref{eq:stab}, which can be explicitly written as:
\begin{align}
\label{eq:Rzero}
R_0  = \frac{1}{\mu \atw} \ \left(1 + \frac{1}{4\atw} (q^2_{\out}-q^2_{\in})\right)^{-1} \ ,
\end{align} 
where we have used the small parameter $\atw$ defined in Eq.~\eqref{eq:alpha}. Notice that this does not depend on the phase difference $t_{0}$. As we will confirm in Sect.~\ref{numerics} in the thin-wall approximation the field inside, $\phiin$, is quite small in any case.

The corresponding action is then simply given by the usual Coleman bounce action,
\begin{align}
\label{eq:action}
 S_{ext} \equiv \frac{27 \pi^2}{2} \frac{\sigma^4}{\vep_0^3} = \frac{\pi^2 \s}{2} R_0^3 \ .
\end{align}

\subsection{True vacuum bubble evolution}\label{sec:membrane}
While bubbles formed from a proper vacuum state always grow this is not necessarily the case when we consider an oscillating vacuum. It is therefore
a key issue when studying tunnelling in an oscillating background to find out when the bubble collapses instead of growing. We present in this section a simple analytical criterium distinguishing the two phases.

Once the bubble has nucleated, its subsequent evolution will be coupled to the evolution of the field inside and outside of it. As in the previous section, we will now consider the coupled evolution of the field inside the bubble (considered homogeneous) and of the bubble wall itself. The field outside of the bubble is considered as a oscillating background. The equation for the bubble radius is given by Eq.~\eqref{Requation},
\begin{align}
\label{eq:rEOM2}  \ddot{R} +2 \frac{1-\dot{R}^2}{R}  &~=~   \frac{p}{\s} (1-\dot{R}^2)^{3/2}
\end{align}

A great deal of simplification can be obtained by averaging over the oscillations of the fields inside and outside the bubble. Let us concentrate on the evolution of the bubble after it nucleates, with $\dot{R} \ll1$. In the thin-wall approximation, we can neglect the slow modulation of time scale $R_0$, so that we have approximately:
\begin{align}
\label{eq:varp}
 \vep & ~\simeq~ \vep_0 = -2  b c ~(1 + \frac{1}{4 \atw} (q_{\out}^2 - q_{\in}^2)) \nn \\
 R  & ~\simeq~  R_0 \nn \\
 p & ~\simeq ~2 b c  \left[ 1 + \frac{q_{\out}^2 \cos 2 \mu t  - q_{\in}^2 \cos 2 \mu (t-t_0) }{4 \atw}   \right] \ .
\end{align}
Focussing on the initial evolution of the bubble averaged over a few oscillations of the fields inside and outside of the bubble, we can neglect the second order terms in $\dot{R}^2$ in~\eqref{eq:rEOM2}. Averaging over a period $2 \pi /\mu$ then leads to
\begin{align}
 \langle \ddot{R}  \rangle \big|_0 = \frac{1}{R_0}  \frac{1-\frac{1}{2 \atw} (q_{\out}^2 - q_{\in}^2)}{ \left( 1+\frac{1}{4 \atw} (q_{\out}^2 - q_{\in}^2) \right) }\ .
\end{align}
Assuming that the subsequent evolution of the bubble will be governed by its behaviour during the first oscillations (an assumption which we will later confirm numerically up to  large oscillation amplitude in Sect.~\ref{sec:lattice}.), it is clear that the bubble will contract if:
\begin{align}
\label{eq:bubblegrowth}
 q_{\out}^2 - q_{\in}^2 > 2 \atw \ .
\end{align}
We can recast Eq.~\eqref{eq:bubblegrowth} as a condition on the initial radius of the bubble, so that all bubbles with radius larger (smaller) than a critical radius $R_c$ will grow (collapse). Using the expression~\eqref{eq:bubblegrowth} one can write:
\begin{align*}
 R_c  = \frac{2}{3 \mu \atw}  \ .
\end{align*}
This last result has in fact a very simple interpretation. Neglecting the field oscillations and expanding $\sqrt{1-\dot{R}^2}$ in $\dot{R}$ from Eq.~\eqref{eq:Hamiltonian} leads to the ``vacuum'' bubble potential:
\begin{align*}
 V_{vac}  ~=~ 4 \pi \sigma R^2 - \frac{4}{3} \pi (2 b c) R^{3} \ .
\end{align*}
This potential has a maximum precisely at $R_c$, so that every bubble with radius larger (smaller) than $R_c$ will grow (contract). Our Eq.~\eqref{eq:bubblegrowth} simply states that while the initial bubble radius depends on the amplitude of the oscillations, the subsequent average evolution of the bubble radius only depends on the asymmetry of the potential. This is in fact consistent with the fact that the evolution of the bubble in~\eqref{eq:rEOM2} is controlled by the differential pressure, for which the effects of the harmonic oscillations cancel in average (cf. Eq.~\eqref{eq:varp}).

 \begin{figure}[t]
	\begin{center}
		\includegraphics[width=0.75\textwidth]{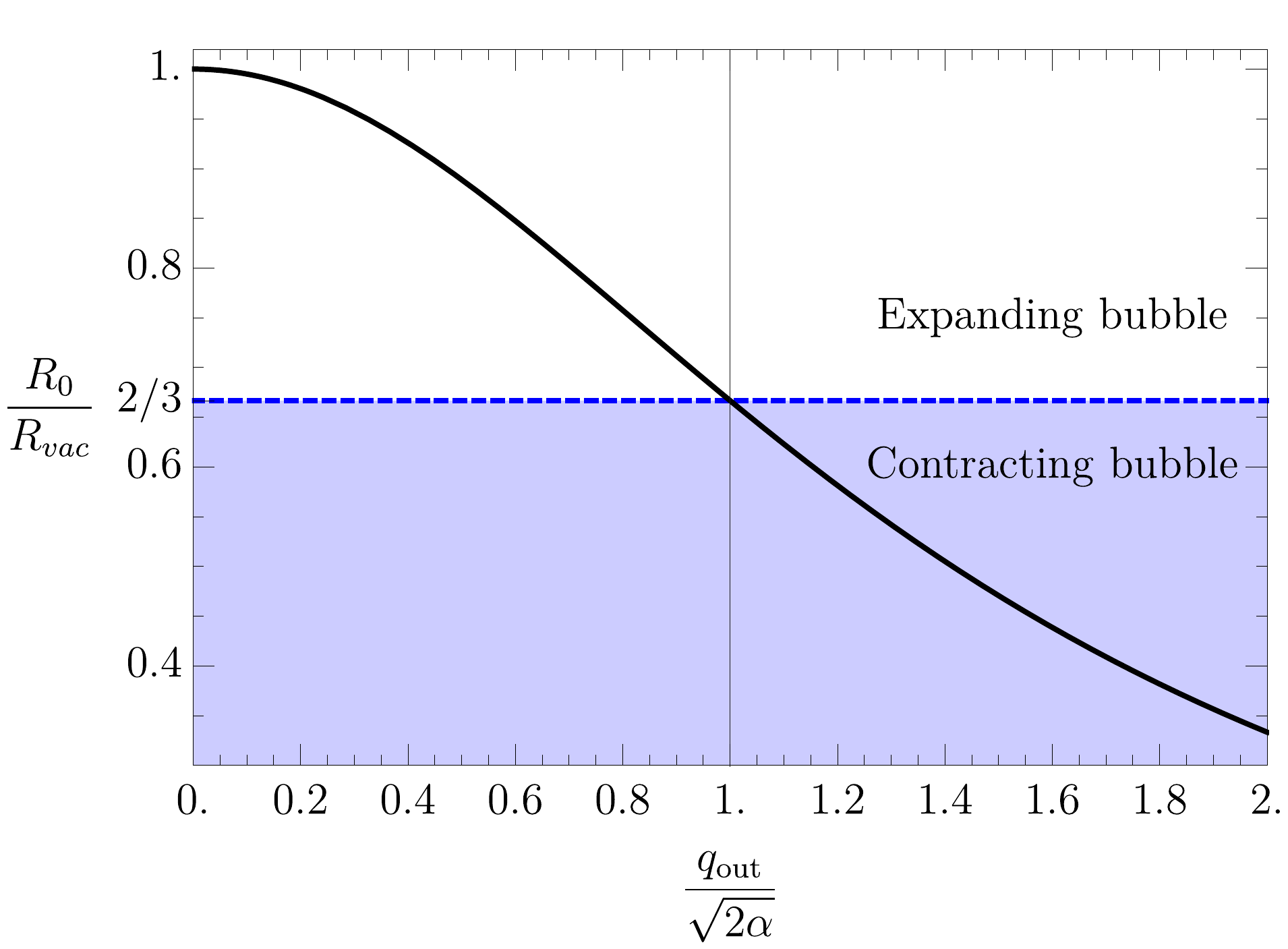}
		\caption{Ratio of the intial radius $R_0$ over the initial radius in absence of oscillation $R_{vac}$ in the case where $q_{\in} = 0$. The dash horizontal line gives the critical radius for bubble growth.}
		\label{fig:ratioR}
	\end{center}
\end{figure}

\begin{figure}[t]
	\begin{center}\subfloat[]{%
\includegraphics[width=0.515\textwidth]{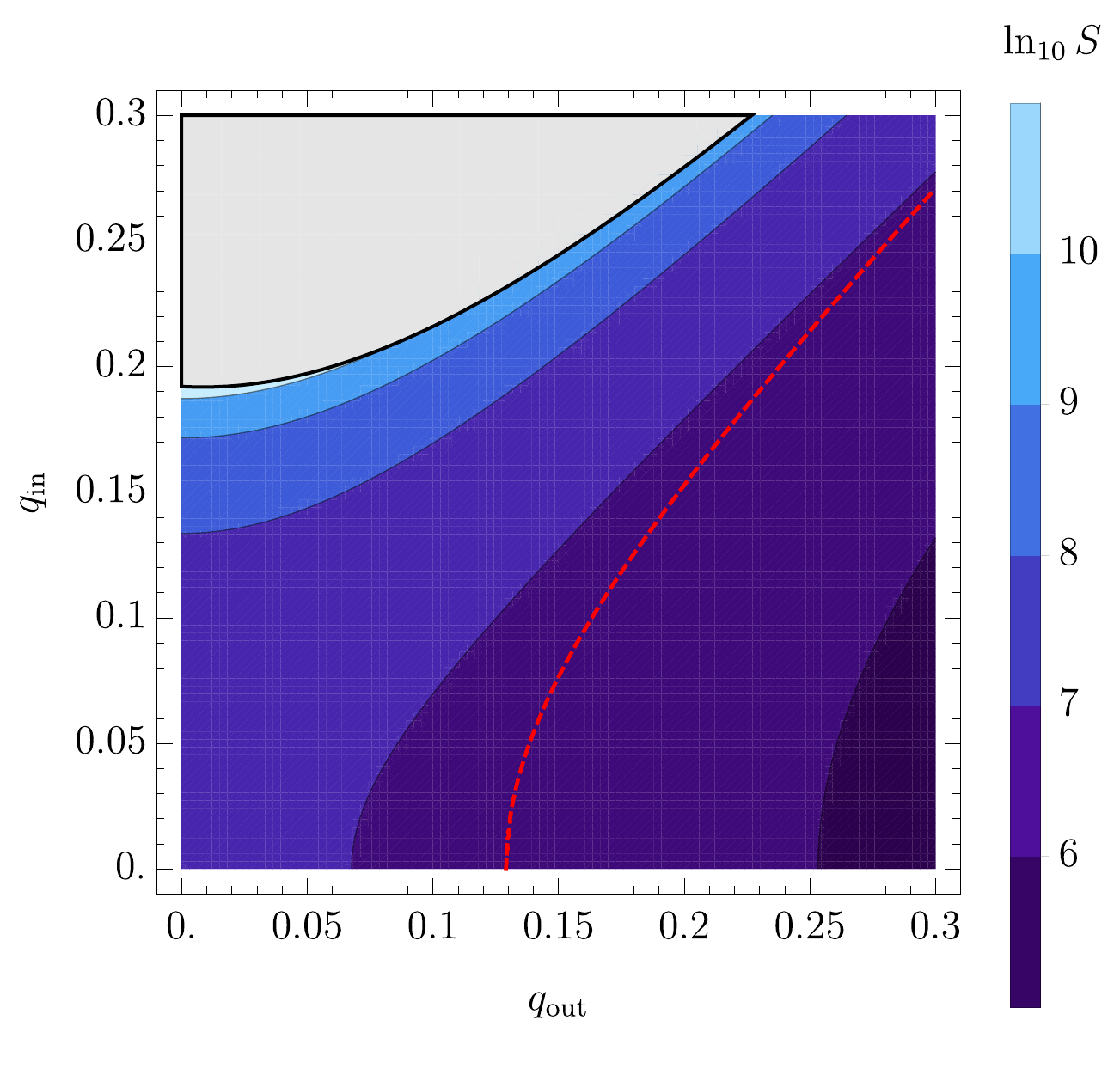}
}%
\hspace{0pt}
\subfloat[]{%
\includegraphics[width=0.455\textwidth]{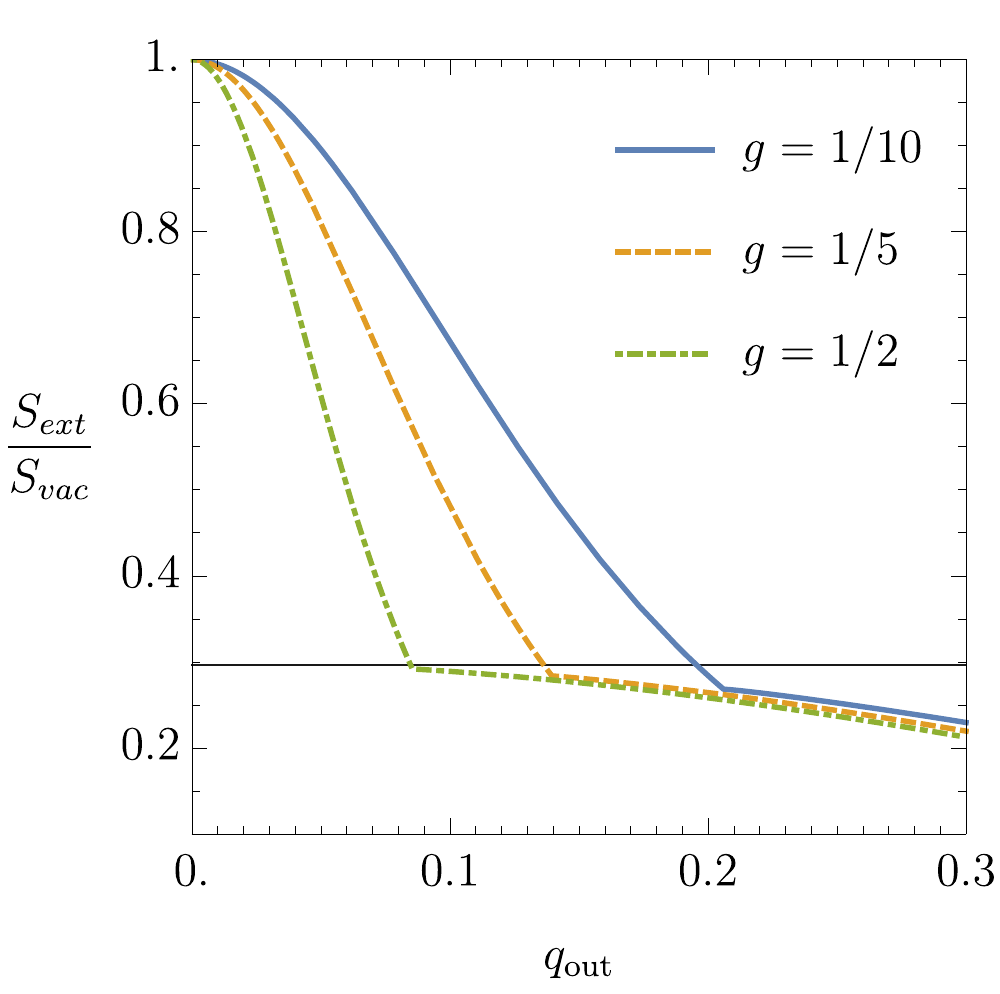}
}%

		\caption{\textbf{(a)}: Coleman bounce action for all the bubbles satisfying energy conservation given an oscillation amplitudes $q_{\in}$ ($q_{\out}$) inside (outside) the bubble. The red dashed line indicates the boundary between growing (above the line) and contracting bubbles. We have used $b=1/300, c=1$ and $g=1/5$. \textbf{(b)}: Ratio of the minimum Coleman bounce action over the vacuum to vacuum one as function of the initial amplitude of oscillation $q_{\out}$ for various values of $g$.}
		\label{fig:phase}
	\end{center}
\end{figure}

We illustrates this behaviour in Figure~\ref{fig:ratioR} by showing the ratio of the initial radius $R_0$ over the initial radius in absence of oscillation $R_{vac}$ in the case where $q_{\in} = 0$. In the more general case with $q_{\in} \neq 0$, we have represented the boundary for bubble growth as the red dashed line in Figure~\ref{fig:phase}a. This figure furthermore gives the usual Coleman bounce action $S_{ext}$ for a tunnelling happening at the extremum of the oscillation of the initial field as expressed in~\eqref{eq:action}.
Note that in principle, the MPEP should associate one $q_{\out}$ to only one $q_{\in}$, and only for this value the result~\eqref{eq:action} is applicable. In particular, we will see that for small harmonic oscillations $q_{\in} \sim 0$ from quite generic arguments. For larger field oscillations, we have in general $q_{\in} > 0$ as we will show in Sect.~\ref{numerical}.

It is interesting to evaluate the minimum action $S_{ext}$ leading to a growing bubble for a given initial oscillating state of amplitude $q_{\out}$. This is straightforwardly derived from~\eqref{eq:Rzero},~\eqref{eq:bubblegrowth} and~\eqref{eq:action}, leading to
\begin{align}
\label{eq:ratioBextr}
 \frac{S_{ext}}{S_{vac}}~=~ \begin{cases} \ \displaystyle \left( 1+ \frac{q_{\out}^2}{4 \atw} \right)^{-3}  \phantom{  \frac{2}{3}}\text{   for $q_{\out} < \sqrt{2 \atw}$}\\[0.9em] \ \displaystyle \left( \frac{2}{3} \right)^3 \phantom{ 1 + \frac{q_{\out}^2}{4 \atw} } \text{  for $q_{\out} > \sqrt{2 \atw}$}\end{cases} \ ,
\end{align}
where the vacuum-to-vacuum action is:
\begin{align}
\label{eq:Svac}
 S_{vac} \equiv \pi^2 \frac{\mu^4 c^5}{3 b^3} \ .
\end{align}

We have represented this interesting saturation behaviour in Figure~\ref{fig:phase}b which show the ratio of the  minimum bounce action allowing for a growing bubble over the vacuum to vacuum one as function of the initial amplitude of oscillation $q_{\out}$. In particular, the saturation is reached when $R_0 \sim R_c$ along the red dashed line of Figure~\ref{fig:phase}a. Notice that we have further included contributions from the anharmonicity of the potential,\footnote{However, we still neglect the change in oscillation frequencies induced by the anharmonicity and treat both oscillating parts as simple harmonic oscillations. A more complete treatment is done numerically in the next section.} which leads to the slow decrease at large $q_{\out}$.

Notice that the results of the previous sections are also applicable in the limit where $ \atw \ll q_{\out}^2 \ll 1$, namely when the oscillations dominates over the asymmetry of the potential.\footnote{In the standard thin-wall approximation the thin-wall parameter $\alpha_{tw}$ compares the energy difference between the two vacua $\sim bc$ to the typical height of the wall $\sim \mu^2c^2$, leading to $\alpha_{tw}\sim b/(\mu^2c)$. If we have an oscillation, the energy difference between the turning point and the true vacuum increases by $\sim q_{\out}^2c^2\mu^2$. Comparing again to the height of the wall we find that $q_{\out}^2\ll1$ should ensure the validity of the thin-wall limit.} However, energy conservation (more precisely the fact that the bubble can form in the first place with a finite radius), implies: $q^2_{\in}-q^2_{\out} < 4\atw$. In this regime it is clear from Eq.~\eqref{eq:bubblegrowth} that the only growing bubbles will be the one with $q_{\in} \sim q_{\out}$. As we will see in the next section, the MPEP leads generally to $q_{\in} \sim 0$, we expect these types of bubble to collapse. This is particularly important for a ``backward'' tunnelling where one supposes that the field is initially oscillating around the true vacuum. Indeed, even starting from the deeper minimum, it is possible to nucleate a bubble of false vacuum satisfying $q^2_{\in}-q^2_{\out} < 4\atw$ if the initial oscillations are larger than the energy difference between the minima (i.e negative $\atw$). However, Eq.~\eqref{eq:bubblegrowth} then implies that such bubble should subsequently collapse so that no phase transition can actually occur.
Note however, that all these considerations above  can be applied directly only within the approximations of this section. Different solutions, for instance including additional spatial perturbations, could in principle still mediate a phase transition in this setup.

\subsection{Time-dependent decay rate}

We now turn to the estimation of the probability for such a bubble of true vacuum to form when the initial field is not at an extremum of its oscillation. Stricto sensu, the usual Coleman bounce result can only be applied when the initial oscillating field $\phiout$ of the form of Eq.~\eqref{eq:fieldsbubble}, $\phiout = q_{\out} \cos \mu t $, is at an extremum of its oscillation. Indeed, one can smoothly rotate to Euclidean time only when the time-derivative of the field vanishes. However, it is clear that given the exponential form of the decay rate, the probability should be dominated precisely by the extremum, since this is where the energy difference between the inside and the outside will be maximised. We will now show that this is what happens in the thin-wall limit.

We will present in a separate, more theoretical paper~\cite{2Dreduction}, a complete description of this process based on the use of the Functional Schr\"odinger equation and of an adaptation of the approach of~\cite{Bowcock:1991dr} of multidimensional tunnelling in quantum mechanics. Let us however use here a more heuristic approach sharing similarities with the calculations and method of~\cite{KeskiVakkuri:1996gn}.\footnote{However, and in contrast with~\cite{KeskiVakkuri:1996gn}, we only rotate the bubble part of the action to Euclidian time, and leave the field outside of the bubble frozen during the tunnelling. Consequently, we disagree with their results regarding tunnelling from an oscillating initial state. In particular, unlike~\cite{KeskiVakkuri:1996gn}, we find that our tunnelling rate tends to be a small multiplicative correction of the tunnelling exponent compared to the standard Coleman result in the vacuum-to-vacuum case and it remains so for all values of the bubble radius. Notice that we have also neglected subdominant corrections from the anharmonicity of the potential compared to~\cite{KeskiVakkuri:1996gn}.}

Our first step is to estimate the amplitude of the oscillations $q_{\in}$ within the bubble after it nucleates. Such oscillations are a perfectly sensible solution of the classical perturbative equation of motion around the bubble solution. However, they are in fact suppressed during the tunnelling process itself. Indeed, suppose that the bubble radius follows its usual evolution in Euclidean time $\tau$ (corresponding to a Coleman $O(4)$-symmetric solution):
\begin{align}
 R (\tau) = \sqrt{R_0^2 - \tau^2} \ ,
\end{align}
for $\tau$ between $-R_0$ and $0$. The Euclidean equivalent of the equation of motion~\eqref{phiinequation} is:
\begin{align}
\label{eq:EoMQEucli}
 \displaystyle {\phiin''} + 3\frac{\phiin' R_0}{R^2} -\mu^2  \phiin = 0\ .
\end{align}
Normalising $\tau$ by $\tilde{\tau} =   \mu \tau$, we can write this equation as:
\begin{align}
  \displaystyle { \phiin''} + 3 \atw \frac{\phiin' (-\tilde{\tau} \atw)}{1-\tilde{\tau}^2 \atw^2 } -  \phiin = 0\ .
\end{align}
Therefore, at zeroth order in $\atw$, the final amplitude of the $ \phiin $ oscillations will be suppressed by $\exp (- \mu R_0 )$. Notice that this depends crucially on the reliability of the thin-wall limit (as we will see from the numerical study of the next section). Overall, we can therefore assume that along the preferred field path during tunnelling, the amplitude of any perturbations inside the bubble should be washed out, so that when the bubble nucleates, we have $q_{\in} \simeq 0$.

Next, we will suppose that during the tunnelling the field oscillations outside of the bubble do not evolve. We therefore neglect corrections to the tension $\sigma$ and suppose that the pressure during the tunnelling is ``frozen'' to its value at the time $t_0$ of bubble nucleation:
\begin{align*}
 p_0 &~= \left(\frac{1}{2} \dot{\varphi}_{\in}^2 - V(c+\phiin)\right) - \left(\frac{1}{2} \dot{\varphi}_{\out}^2 - V(-c+\phiout) \right) \bigg|_{t=t_0} \\  
 &~=  2 B c - \frac{\mu^2 c^2}{2} q_{\out}^2 \sin^2 \! \mu t_0 +   V(-c+\phiout) \\
 &~= 2 B c~( 1  + \frac{q_{\out}^2}{4 \atw} \cos  \! 2 \mu t_0 ) \ .
\end{align*}

We then integrate the Euclidean action along the path corresponding to a tunnelling from an extremum of the oscillation: $R(\tau) = \sqrt{R_0^2 - \tau^2}$ where $R_0$ is given in Eq.~\eqref{eq:Rzero}.
From the Lagrangian~\eqref{eq:Lmembrane} and keeping only the bubble part, we obtain
\begin{align*}
 U &= i \int_{-R_0}^0 d\tau \left[ -4 \pi \s R^2 \sqrt{1+R'{}^2} + \frac{4}{3} \pi p_0 R^3  \right] \\
 &= i 4 \pi \s  \int_{-R_0}^0 d\tau \left[ - R R_0 + \frac{1}{3} \frac{p_0}{\sigma} R^3  \right] \ ,
\end{align*}
and after a bit of algebra we have
\begin{align}
\label{eq:EucliBounce}
 U =  - i \frac{\pi^2 \sigma}{4} R_0^3 \left( 1 + \frac{3 q_{\out}^2}{2\atw}\sin^2 \! \mu t_0 \right)  \ .
\end{align}
In the limit $q_{\out} = 0$ we recover the usual result from Coleman as $S = 2 i U$. 
In particular, we recover the result of Eq.~\eqref{eq:ratioBextr} for $t_0=0$. 

The result found in~\eqref{eq:EucliBounce} has the surprising feature that the decay rate for tunnelling from a field value smaller than $\phifv$ is the symmetric of the one for a field value larger than $\phifv$ (one should expect tunnelling from field value smaller than $\phifv$ to be suppressed). This property is a consequence of the fact that we have neglected the variations in the tension $\sigma$ stemming from the oscillations. This issue is irrelevant for our current purpose, since it can at most lead to a factor two change in the tunnelling rate, hence negligible with respect to the exponential exponent we are estimating. It would still be interesting to quantify precisely the effect of this region, but this should come hand-to-hand with an estimation of the quantum corrections to the tunnelling rate since they are expected to be of the same order\footnote{Especially since it is not clear that one can find a MPEP from the true vacuum to a field value smaller than $\phifv$, since the Euclidean Lagrangian is not bounded in this direction. }.

Another comment is that the wall region is now an interface between a quantum and a classical region. While the corrections to the tension from the oscillations during the classical evolution were found to be subdominant in Sect.~\ref{sec:membrane}, we leave to future work a proper analysis of this region. Moreover, the bubble radius evolution $R(\tau)$ that we used is not the optimal one since we simply used the MPEP from an extremum of an oscillation. Strictly speaking, we therefore expect the tunnelling rate obtained from Eq.~\eqref{eq:EucliBounce} to be only a lower bound. The complete calculation, including the estimation of the optimal path, will be done in~\cite{2Dreduction}, since, as we will now see, this has no effect on the exponential exponent itself in the thin-wall limit.

\begin{figure}[t]
	\begin{center}
		\includegraphics[width=0.7\textwidth]{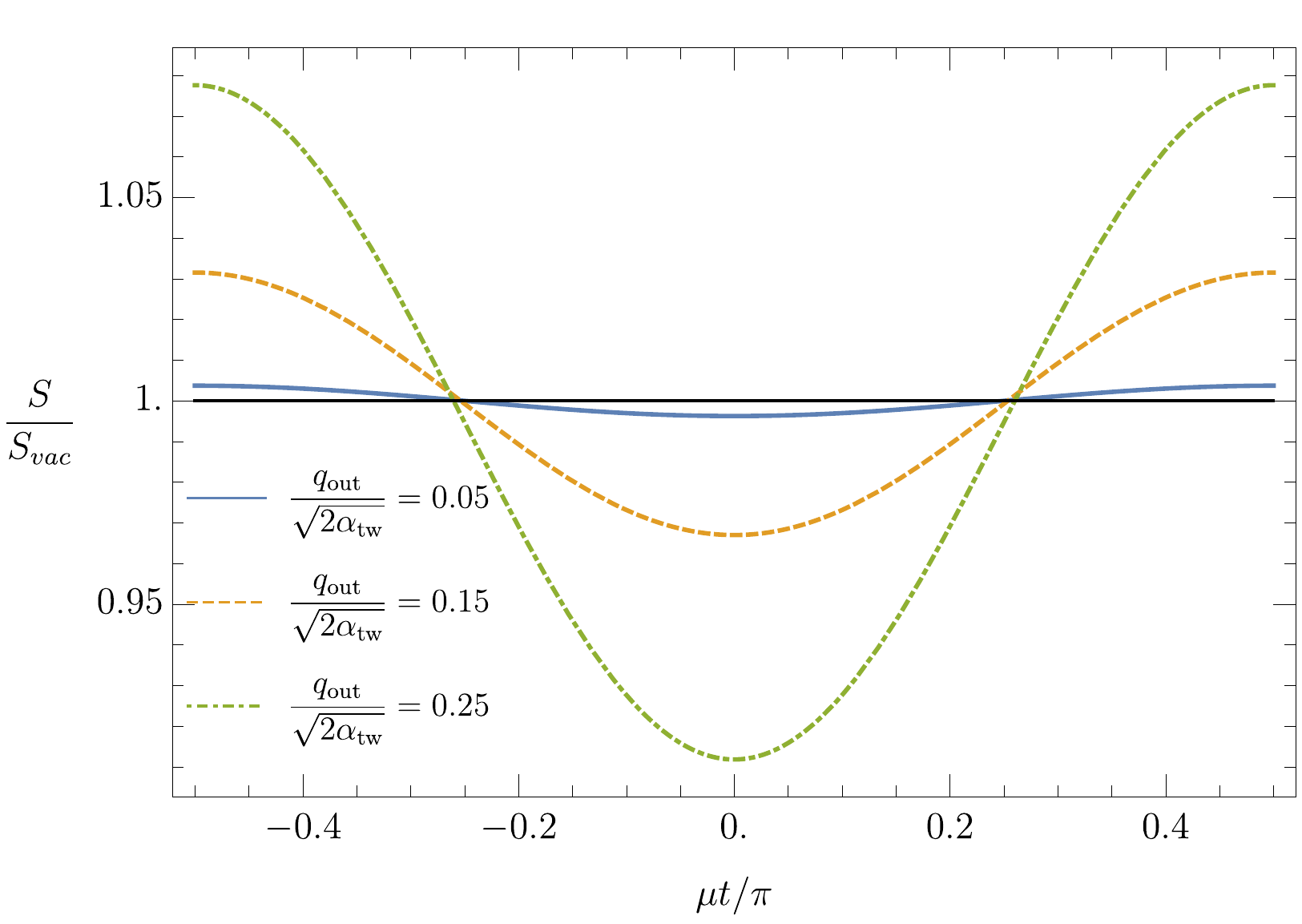}
		\caption{Ratio of the decay rate exponential exponent $S$ over the vacuum-to-vacuum one $S_{vac}$ as function of $\mu t/\pi$ for various values of $q_{\out}/\atw$. The extremum of the oscillation occurs at $t=0$.}
		\label{fig:ratioB}
	\end{center}
\end{figure}

Using the extremum bounce action $S_{ext}$ defined in Eq.~\eqref{eq:action} we can finally write the time-dependent final tunnelling rate $\Gamma$ as:
\begin{align}
\label{eq:ratetime}
 \Gamma (t) \propto \exp \left[\displaystyle -  S_{ext} \displaystyle \left( 1+\frac{3}{2} \frac{q_{\out}^2}{\atw}\sin^2 \! \mu t \right)  \right] \ .
\end{align}
We have illustrated this result in Figure~\ref{fig:ratioB}.  Let us point out that since we are considering large exponential term, even a percent level correction can lead to a large suppression of the final tunnelling rate.

Alternatively, if the field oscillations are fast compared to all other time relevant time scale (e.g. the Hubble rate), we can integrate over a period $\mu^{-1}$ to find:
\begin{align}
\label{eq:rateav}
\big\langle  ~\Gamma (t)~ \big\rangle &~=~ \exp \left[ - S_{ext}(1+\frac{3}{4}\frac{q_{\out}^2}{\atw}) \right] I_0 \left(  S_{ext} \frac{3}{4}\frac{q_{\out}^2}{\atw} \right) \nn \\
&~\simeq~ \frac{1}{\sqrt{S_{ext}} q_{\out}^2/\atw} \exp \left[- S_{ext} \right]\ ,
\end{align} 
Where we have replaced the modified Bessel function by its exponential expansion at large parameter. As could have been expected, the exponential part of the decay probability is dominated by the probability at the extremum of the oscillation, while the rest of the oscillations only produce a non-exponential correction (which should be treated along with the quantum corrections since it is of the same order).

Finally, while we have estimated Eq.~\eqref{eq:ratetime} in the limit of small $q^2_{\out}/\atw$. It is clear that the fact that the tunnelling rate at the extremum of the oscillation dominates the average rate should only be enhanced for larger oscillations. Therefore, as long as one remains in the thin-wall limit, we expect that the final result~\eqref{eq:rateav} should hold.

\section{Numerics}\label{numerics}

All our results up to now are based on the thin-wall approximation and other simplifications necessary to obtain analytical results.
In this section we will present a numerical approach to the bounce from oscillating initial field states and confront our analytical results with these direct numerical calculations.
\subsection{The oscillating bounce}
\label{numerical}
First we will review the standard numerical procedure for finding solutions to our original equation of motion~\eqref{eq:EOM0}. In case of tunnelling from a homogeneous configuration in the false vacuum $\phifv$ the standard boundary conditions needed to obtain a finite action~\eqref{eq:action0} are:
\begin{equation}\label{eq:EOMboundary0}
\dot{\phi}(r=0)=0, \quad \quad \phi(r) \xrightarrow[]{r\rightarrow\infty} \phifv.
\end{equation} 
The simplest method to find a solution obeying these conditions is to iterate different $\phi(r=0)$ always keeping $\dot{\phi}(r=0)=0$ using an overshoot/undershoot algorithm until we find a solution fulfilling the second boundary condition.

Tunnelling from an oscillating state requires different boundary conditions as the second condition in~\eqref{eq:EOMboundary0} cannot be fulfilled if our background is not the field residing in the minimum of the potential.
The most straightforward way to find a bubble profile reaching only to a certain field value $\phifv+\phiout$ is to use what we would previously call undershoot solutions, that is solutions in which the field derivative vanishes before the field reaches the false vacuum.
We thus replace~\eqref{eq:EOMboundary0} with 
\begin{equation}\label{eq:EOMboundary1}
\dot{\phi}(r=0)=0,\quad \dot{\phi}(r=r_{\rm end})=0,\quad \phi(r=r_{\rm end})=\phifv+\phiout,
\end{equation} 
and again find the desired solution using an overshoot/undershoot iterating $\phi(r=0)$ always keeping $\dot{\phi}(r=0)=0$ until the remaining conditions are fulfilled.
Then our initial field configuration is simply the numerically obtained $\phi(r)$ for $r<r_{\rm end}$  and $\phi(r)=\phifv+\phiout$ for $r>r_{\rm end}$. Lastly, we define the numerical initial bubble size $R^{\rm N}_{0}$ as the value for which the field equals $\phi(r_{0})=(\phi(0)-(\phifv+\phiout))/2$.

\begin{figure}
\includegraphics[height=5.2cm]{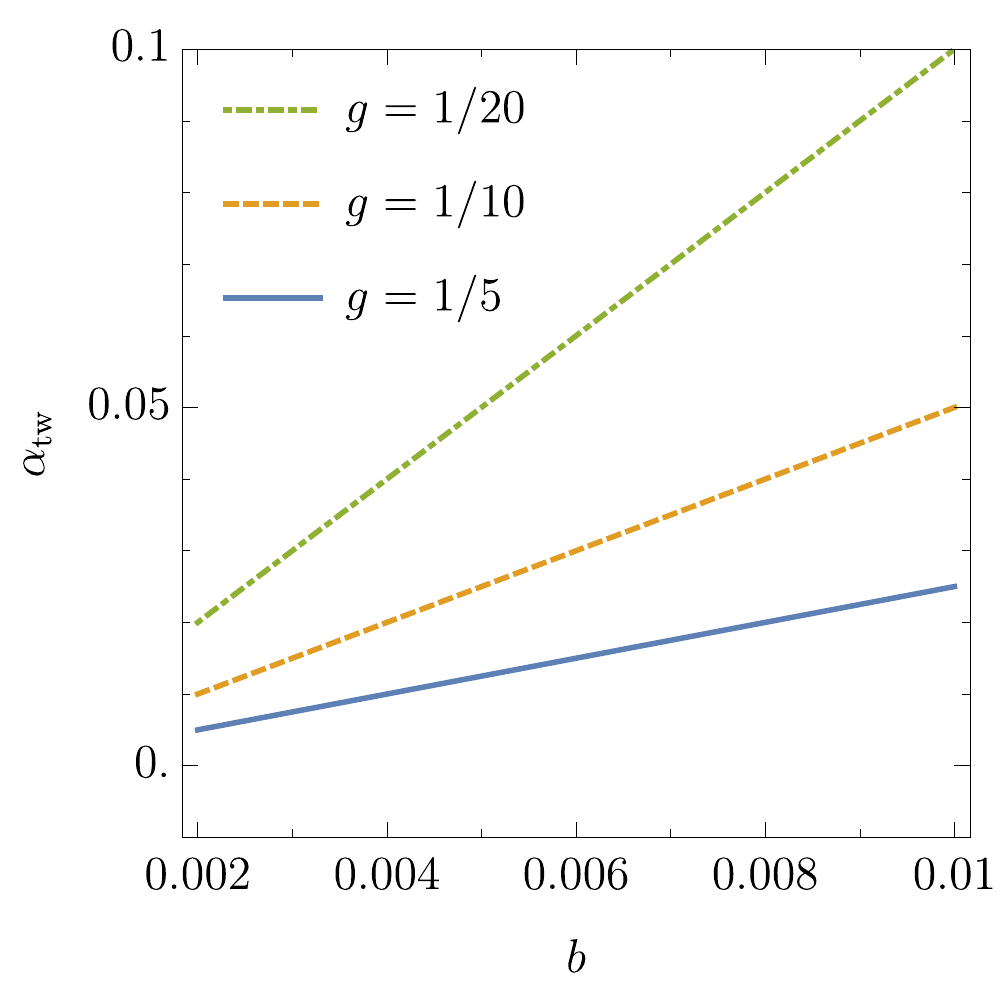}
\includegraphics[height=5.2cm]{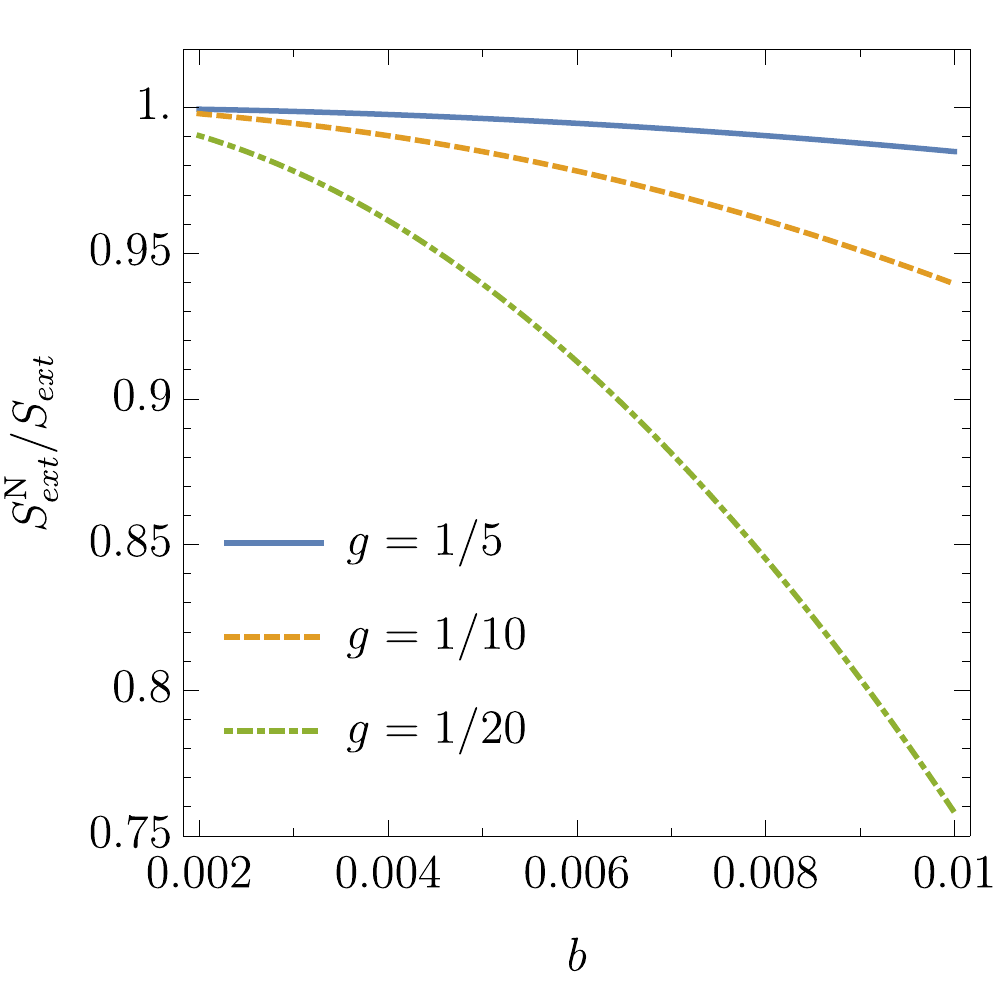}
\includegraphics[height=5.2cm]{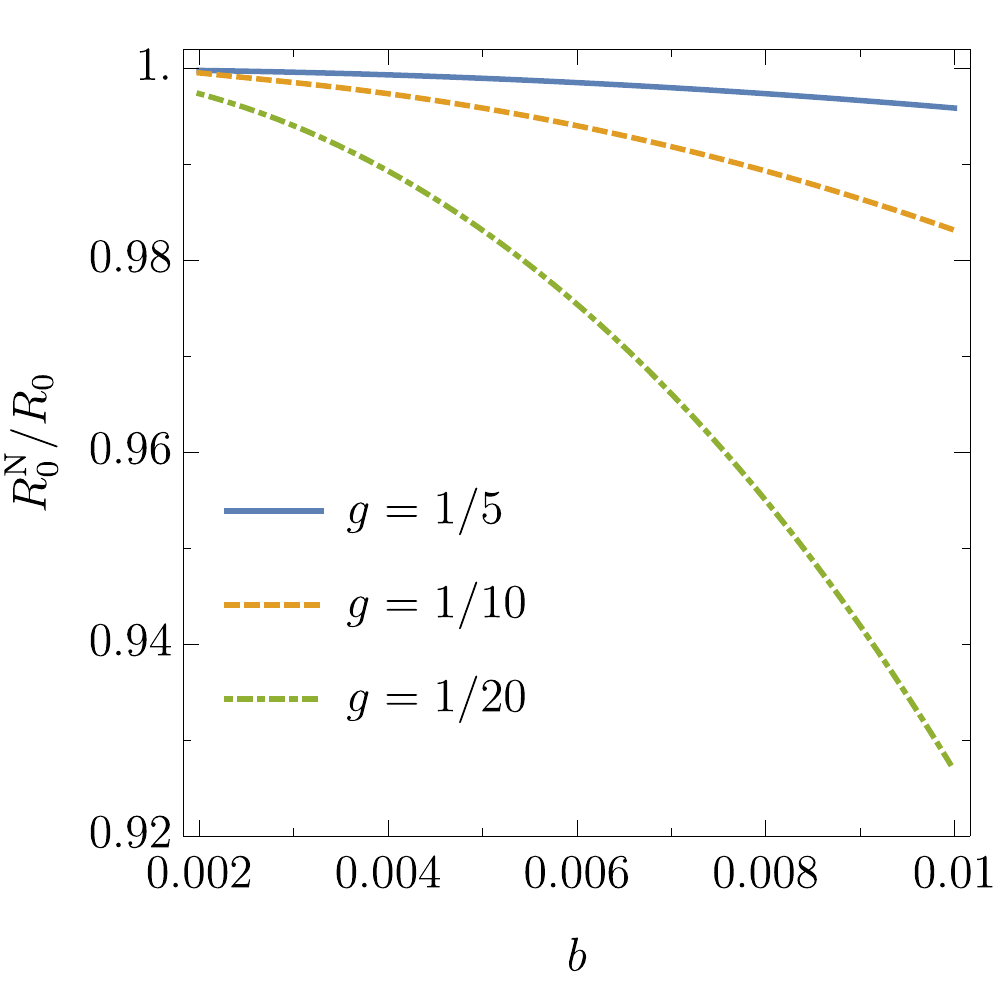}
\caption{
Ratio of the numerically obtained action $S^{\rm N}_{ext}$ and bubble size $R^{\rm N}_{0}$ and the thin-wall results  for these quantities $S_{ext}$ and $R_0$ for a range of parameters $g$ and $b$, assuming no oscillation $\phiout=0$ and setting $c=1$. Going to the extreme thin-wall limit is numerically challenging, this is why we do not show the region $\alpha_{tw},b\to 0$.
\label{fig:TWagreement}}
\end{figure}

Before proceeding to the lattice evolution we will discuss the agreement between our numerical solution discussed above and the thin-wall results from section~\ref{sec:membrane}. Figure~\ref{fig:TWagreement} shows the difference between numerically obtained action $S^{\rm N}_{ext}$ and bubble size $R^{\rm N}_{0}$ and the thin-wall results  for these quantities $S_{ext}$ and $R_0$ for a range of parameters with no oscillation $\phiout=0$. We also plot the corresponding value of the parameter $\atw$ from~\eqref{eq:alpha} which should be sufficiently small for the thin-wall approximation to work.
Just as expected the two methods agree very well as long as the splitting between the vacua controlled by $b$ is sufficiently small.

\begin{figure}
\includegraphics[height=7.5cm]{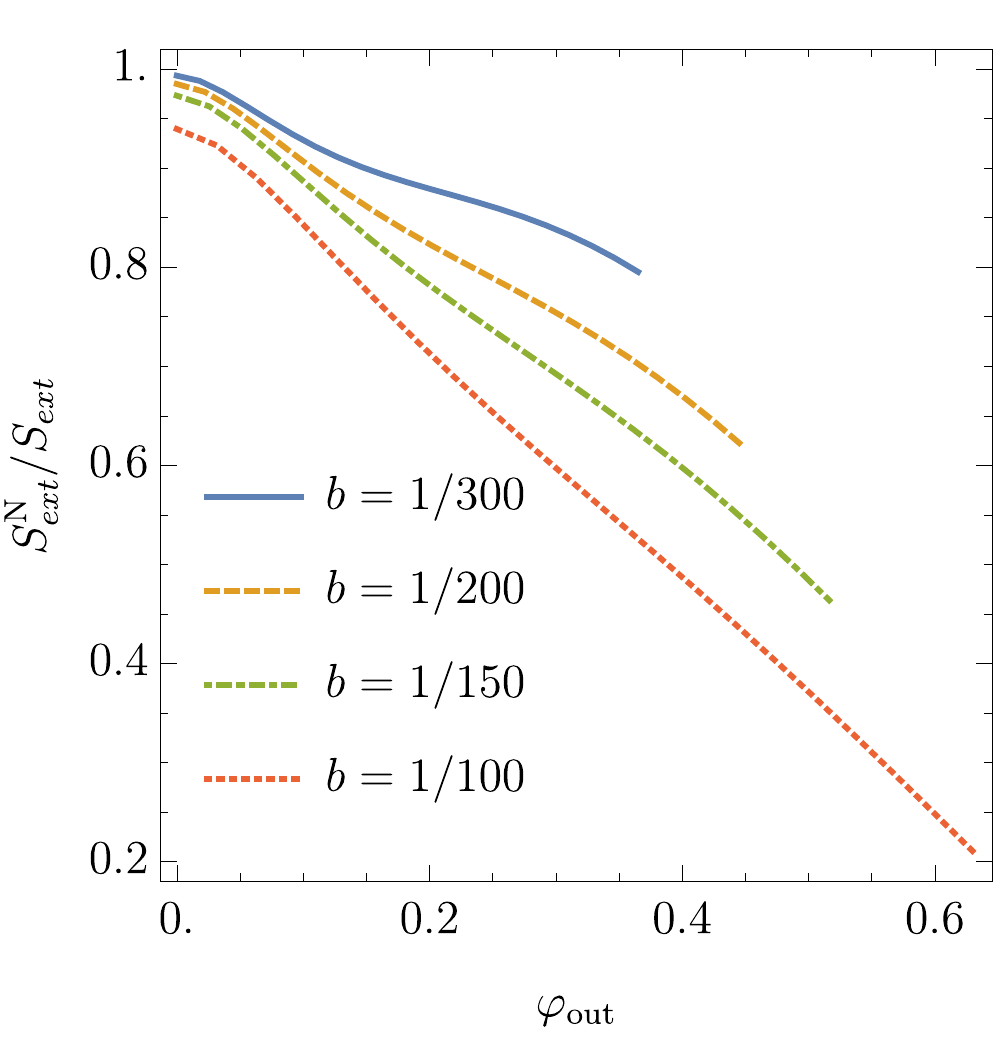}\hspace{1.cm}
\includegraphics[height=7.5cm]{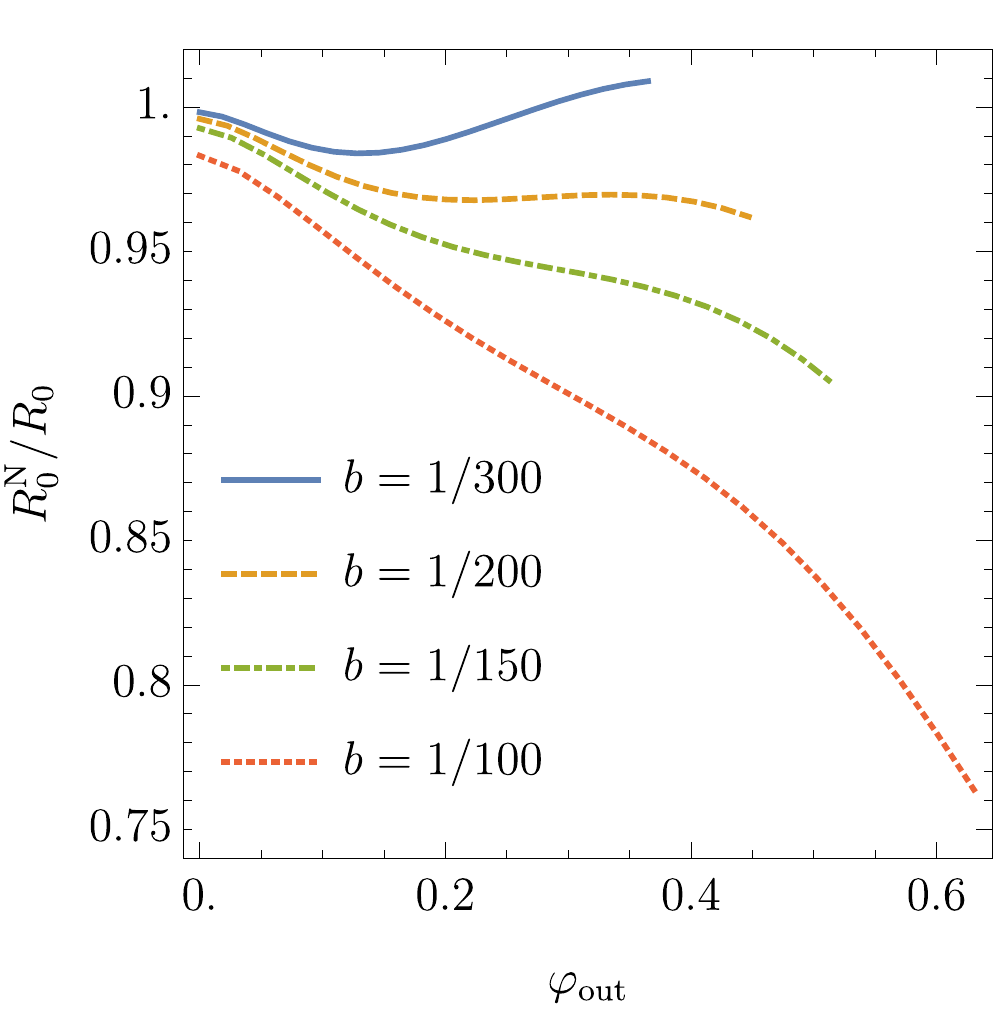}
\caption{
Ratio of the numerically obtained action $S^{\rm N}_{ext}$ and bubble size $R^{\rm N}_0$ and the thin-wall results  for these quantities $S_{ext}$ and $R_0$ for $g=1/10$ and $c=1$ as a function of $\varphi_{\rm out}$.
\label{fig:TWagreement2}}
\end{figure}

Next we will check how the thin-wall approximation holds when the amplitude of oscillation $\phiout$ increases.
 Figure~\ref{fig:TWagreement2} shows the difference between numerically obtained action $S^{\rm N}_{ext}$ and bubble size $R^{\rm N}_0$ and the thin-wall results  for these quantities $S_{ext}$ and $R_0$ for $g=1/10,\, c=1$ and several values of $b$ as a function of the amplitude of the oscillation $\phiout$. We continue our plots up to the value $\phiout= c \sqrt{2\atw}=c \sqrt{b/(\mu^2 c)}$ for each set of parameters, where the analytical approximation should hold.

\begin{figure}
\includegraphics[height=7.5cm]{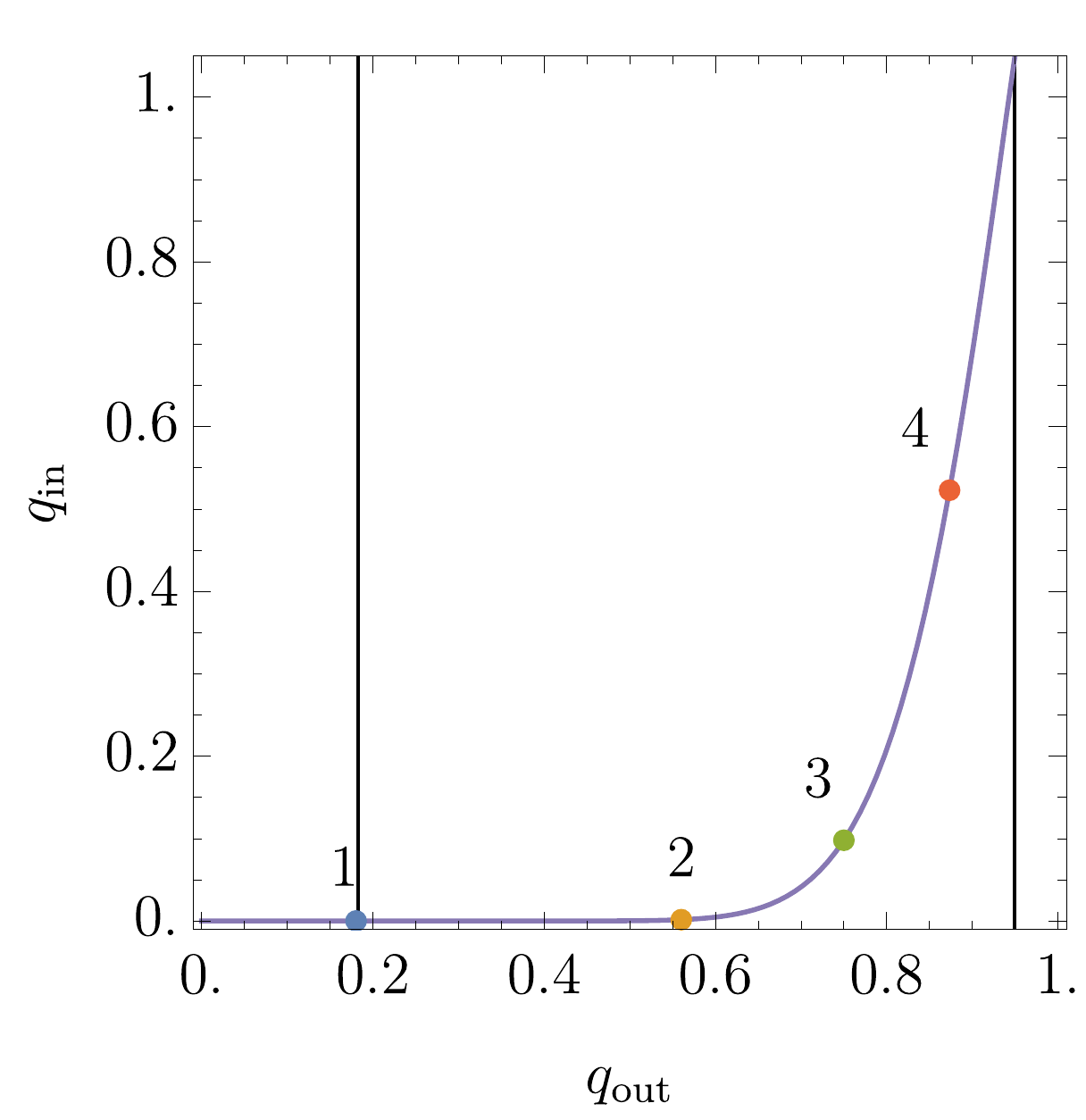}
\includegraphics[height=7.5cm]{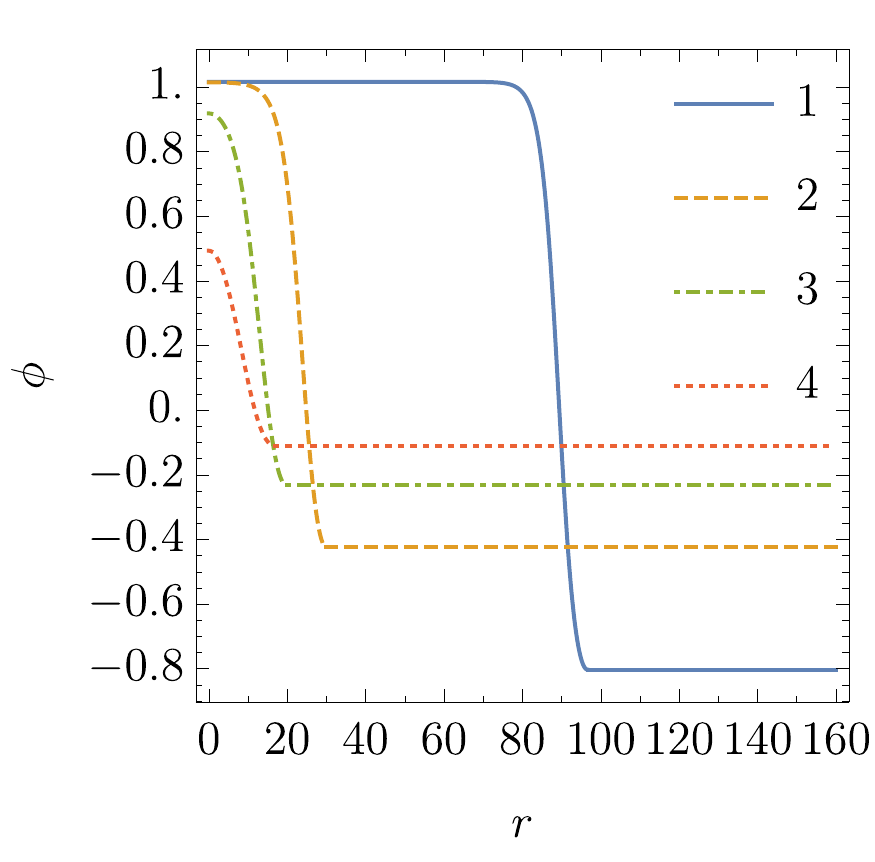}\hspace{1.cm}
\caption{
Value of the field inside the bubble as a function of the field outside for $g=1/10$, $b=1/300$ and $c=1$ . Vertical lines show $q_{\rm out}=\sqrt{2\alpha_{\rm tw}}$ and the field value corresponding to the top of the potential barrier (from left to right). Numbered points correspond to example bubble profiles on the right panel.
\label{fig:phiin}}
\end{figure}

As mentioned in Section~\ref{sec:membrane} the value of the field inside the bubble $\varphi_{\in}$ is an output of a complete algorithm for calculating the bubble profile. As expected for small oscillations this value is very close to the true vacuum and differs from it only for very large oscillations. This is shown in Figure~\ref{fig:phiin} where we also plot several examples of a resulting bubble profile.   

Our main conclusion here is that the numerical results and analytical approximation of the bubble radius agree very well even in the presence of oscillations as long as we are in the thin-wall regime.
The action of our numerical solution is always smaller than the analytical result and the agreement gets worse faster than the bubble size. However within thin-wall limit the results again are in reasonably good agreement.

\subsection{Lattice evolution}
\label{sec:lattice}

\begin{figure}
\includegraphics[height=7.5cm]{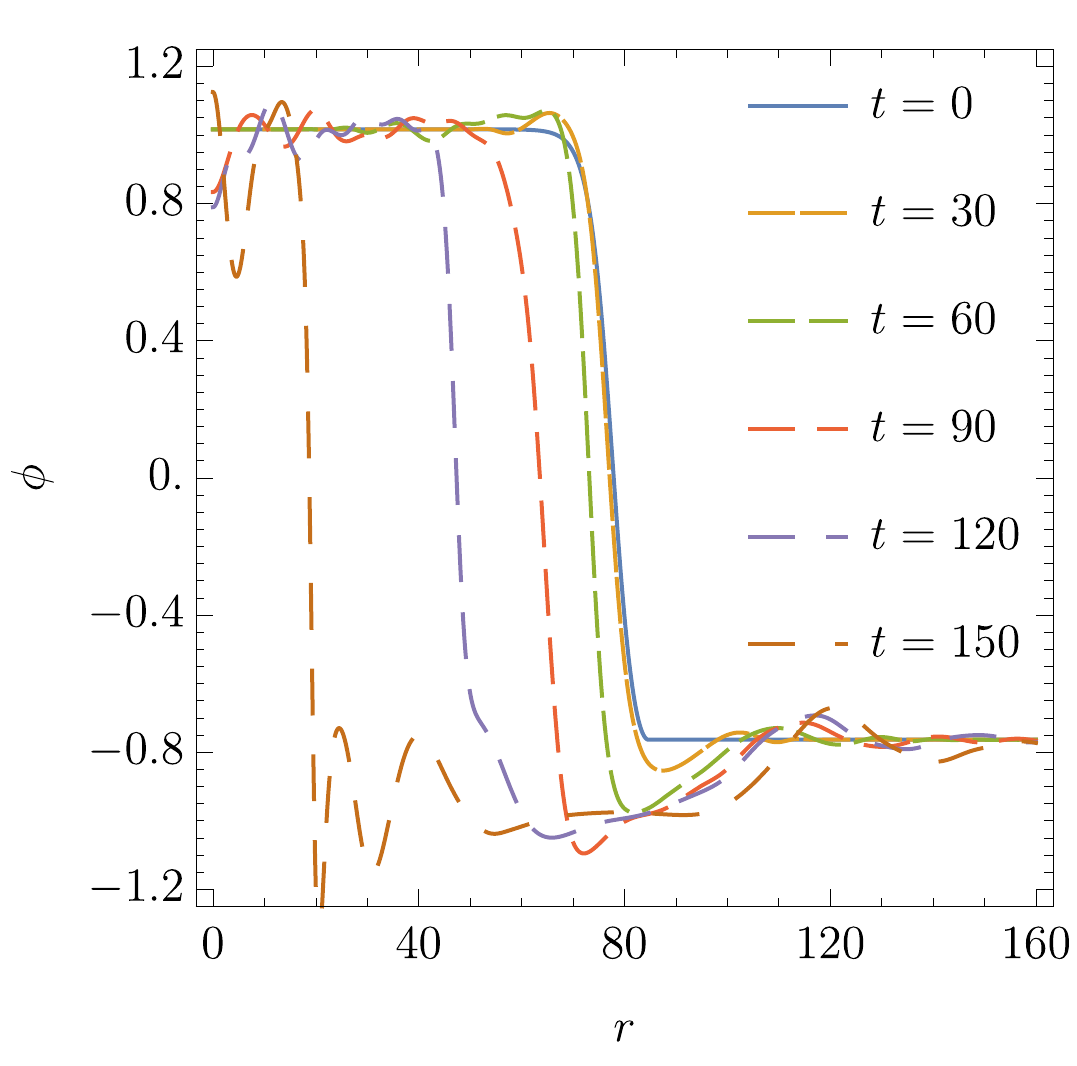}\hspace{0.5cm}
\includegraphics[height=7.5cm]{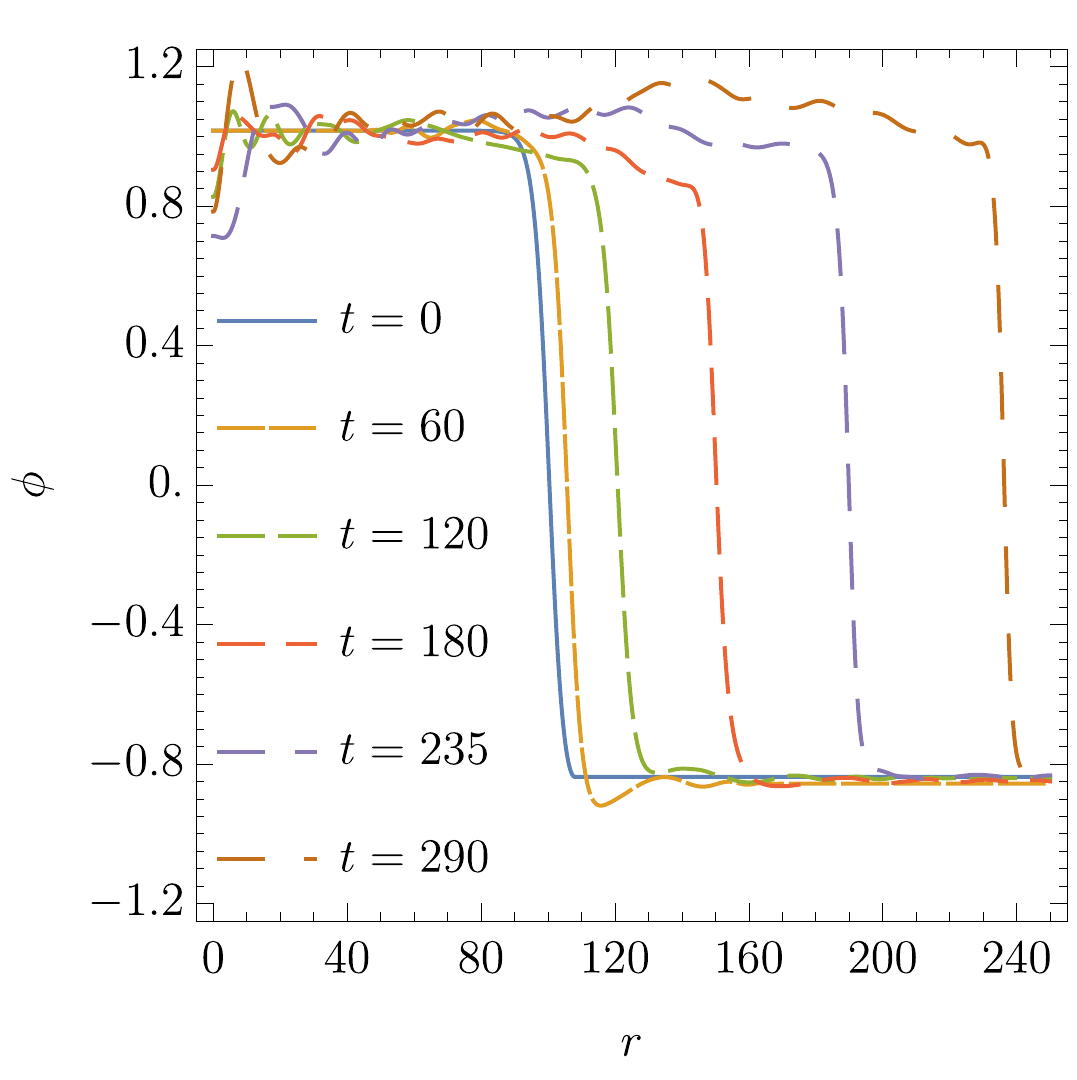}
\caption{
Field profiles showing lattice bubble evolution for oscillation reaching $1.2\sqrt{2 \alpha_{\rm tw}}$ (left panel) and $0.8\sqrt{2 \alpha_{\rm tw}}$ (right panel).
The values defining the potential were set to $g=1/10$, $b=1/300$ and $c=1$.
\label{fig:bubbleevo}}
\end{figure}

Our next step is the verification of the bubble evolution after its nucleation. In Section~\ref{sec:membrane} we presented results based on the membrane approach and now we will verify these by performing a direct lattice calculation. We will as usual assume spherical symmetry of the system 
$\phi=\phi(t,\rho)$ where $t$ is time and $\rho$ is the three dimensional radial coordinate. This leads to the action  
\begin{align}\label{eq:real_action}
 S_R = \int d^4 x
  \left[ \frac{1}{2} \d_\mu \phi \d^\mu \phi - V(\phi)\right]= 4\pi \int d t \rho^2 d \rho
  \left[ \frac{1}{2} (\d_t \phi) ^2 - \frac{1}{2} (\d_\rho \phi) ^2 -
  V(\phi)\right]\ ,
\end{align}
 and the real time equation of motion
\begin{align}\label{eq:real_EOM}
 \frac{\d^2 \phi}{\d t^2}-\frac{\d^2 \phi}{\d \rho^2} -  \frac{2}{\rho} \frac{\d \phi}{\d \rho} =- V'(\phi) \ .
\end{align}
We perform a simple two dimensional lattice evolution treating the bubble profiles discussed in previous subsection as an initial condition. We also set the initial time derivative to a certain value identical for every $\rho$ which means that our bubble simply moves uniformly with the background after nucleation. This fixes our boundary conditions to\footnote{Except at the turning point this initial condition differs from the one we employed in the thin-wall approximation described in the previous section. There we effectively assumed a vanishing time derivative inside the bubble.}
\begin{equation}\label{eq:latticeboundary}
\d_\rho \phi (t,\rho=0)=0, \quad \phi(0,\rho)=\phi_{\rm bounce}, \quad \d_t \phi(0,\rho)=\dot{\phi}_0.
\end{equation} 
After the evolution we calculate the bubble size as a function of time and find the average value normalised to the initial size $\overline{R/R(t=0)}$. If this value is greater than one the bubble grows, while for smaller values the bubble eventually collapses. We show an example of bubble expansion and collapse in Figure~\ref{fig:bubbleevo}.
For bubbles slightly smaller than the critical value we recreate the oscillon solutions discussed in  \cite{Copeland:1995fq} which oscillate for a time before eventually decaying. In general we find very good agreement with the analytical prediction that bubbles will grow as long as their radius is bigger than $2/3$ of the vacuum-to-vacuum bubble radius (which corresponds to the oscillation reaching $\sqrt{2\alpha_{\rm tw}}$). This is not surprising as the bubble obtained numerically and analytically is very similar in the thin wall limit as we show in Figure~\ref{fig:TWagreement2}.  
\begin{figure}
\includegraphics[height=7.5cm]{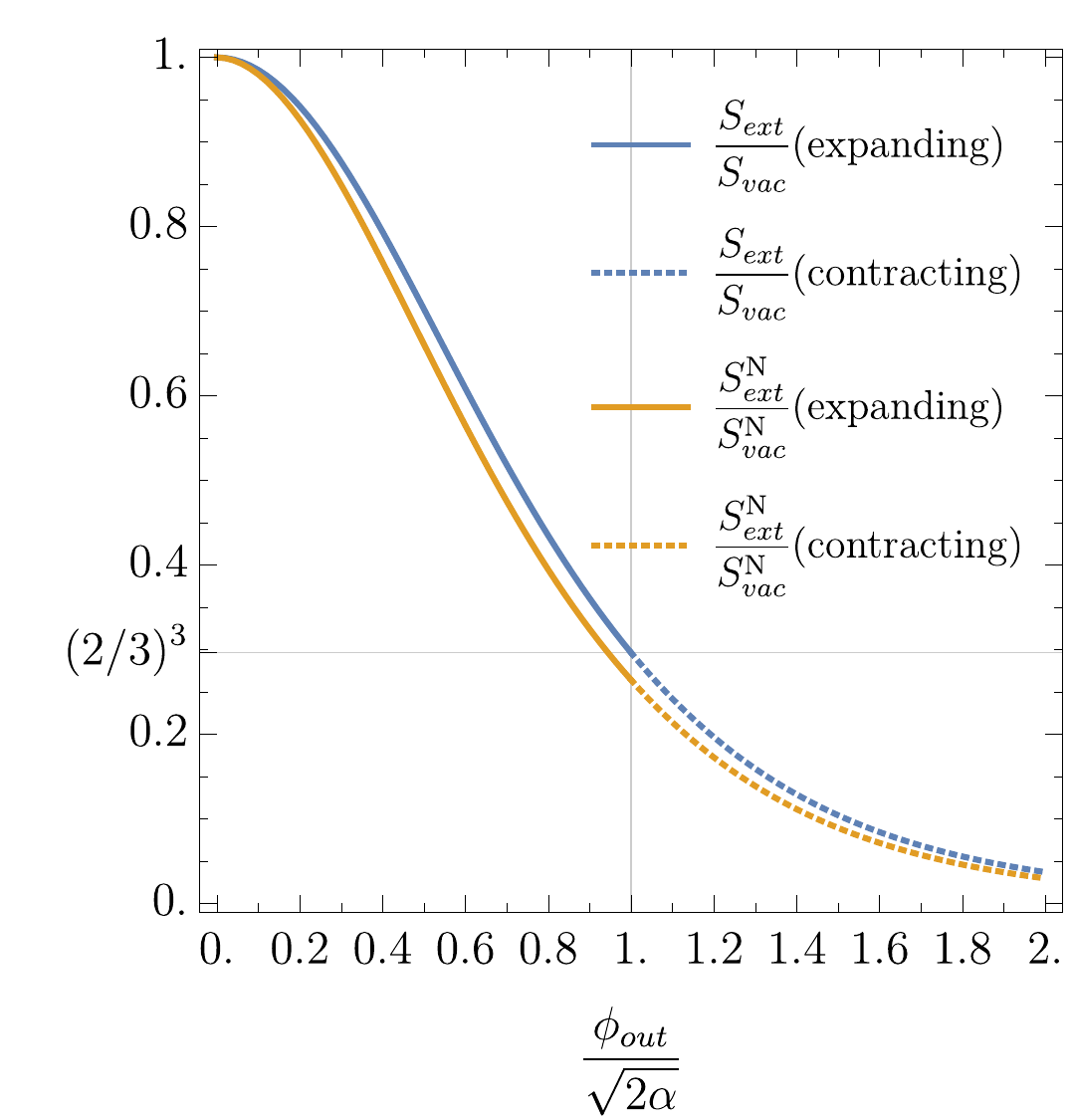}\hspace{1.cm}
\includegraphics[height=7.5cm]{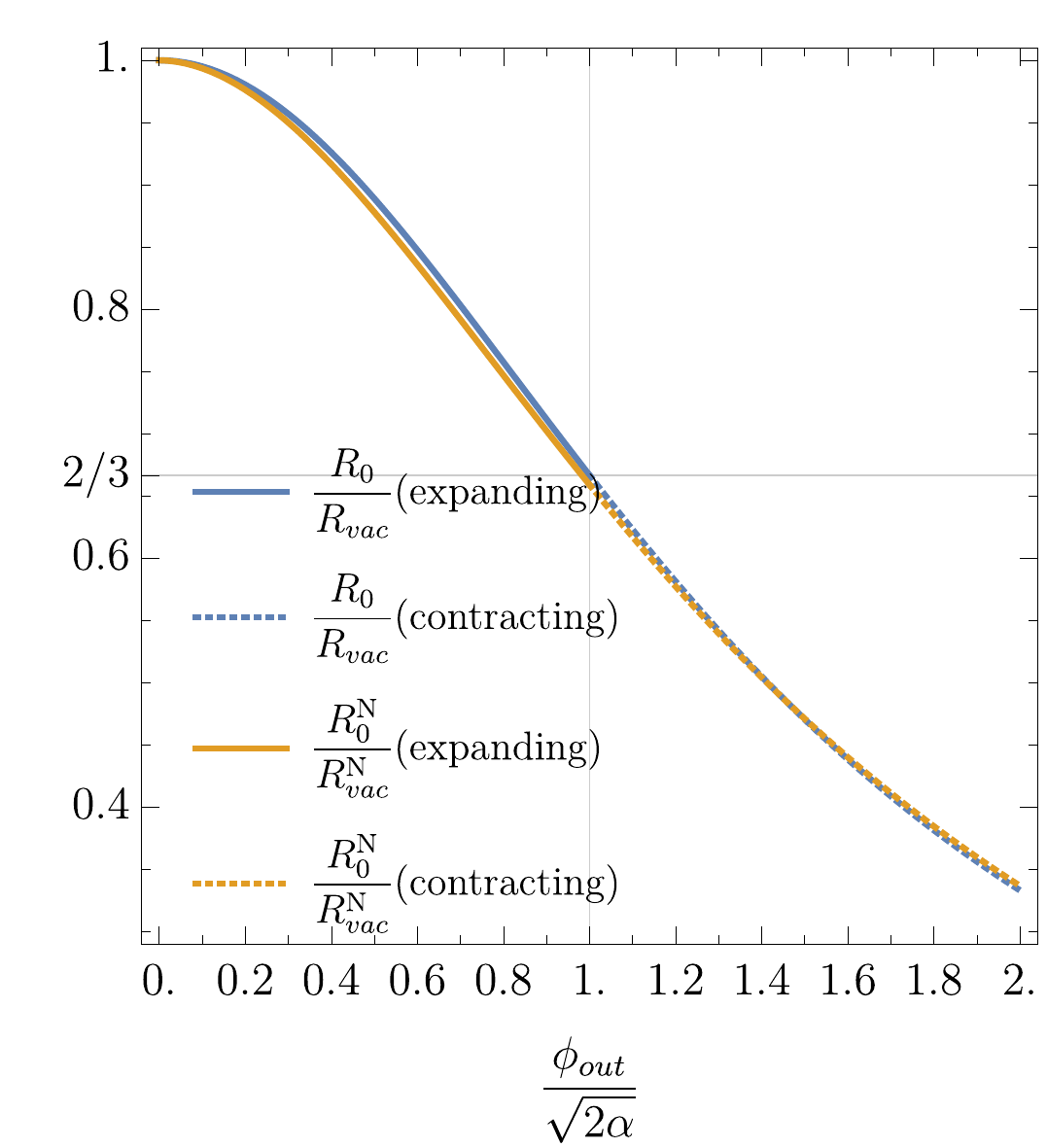}
\caption{
Ratio of the action $S_{ext}$ and bubble radius $R_0$ to the same quantities calculated in absence of oscillation $S_{vac}$ and $R_{vac}$ shown for both analytical results and numerical ones (with subscript $N$).
The values defining the potential were set to $g=1/10$, $b=1/300$ and $c=1$. The dotted line show the region in which the resulting bubble contracts after nucleation for both analytical result and the result of lattice evolution (with superscript $N$).
\label{fig:bubblegrowth}}
\end{figure}

We are finally in the position to average our results over a full oscillation period and compute the resulting decay rate. Numerically, we simply solve the background field equation of motion to get the initial position and corresponding speed within one period of oscillation even for large anharmonic oscillations. We then use the resulting value in our boundary conditions~\eqref{eq:EOMboundary1} to find the numerical bounce and consequently the bounce action. Averaging then gives results very similar to the analytical result in Eq.~\eqref{eq:rateav}. This is not surprising as the final rate is basically controlled by the action of the smallest growing bubble. The main difference between the analytical prediction for a growing bubble and our lattice results comes from the initial speed of the background in which the bubble nucleates. 

\begin{figure}
\includegraphics[height=7.5cm]{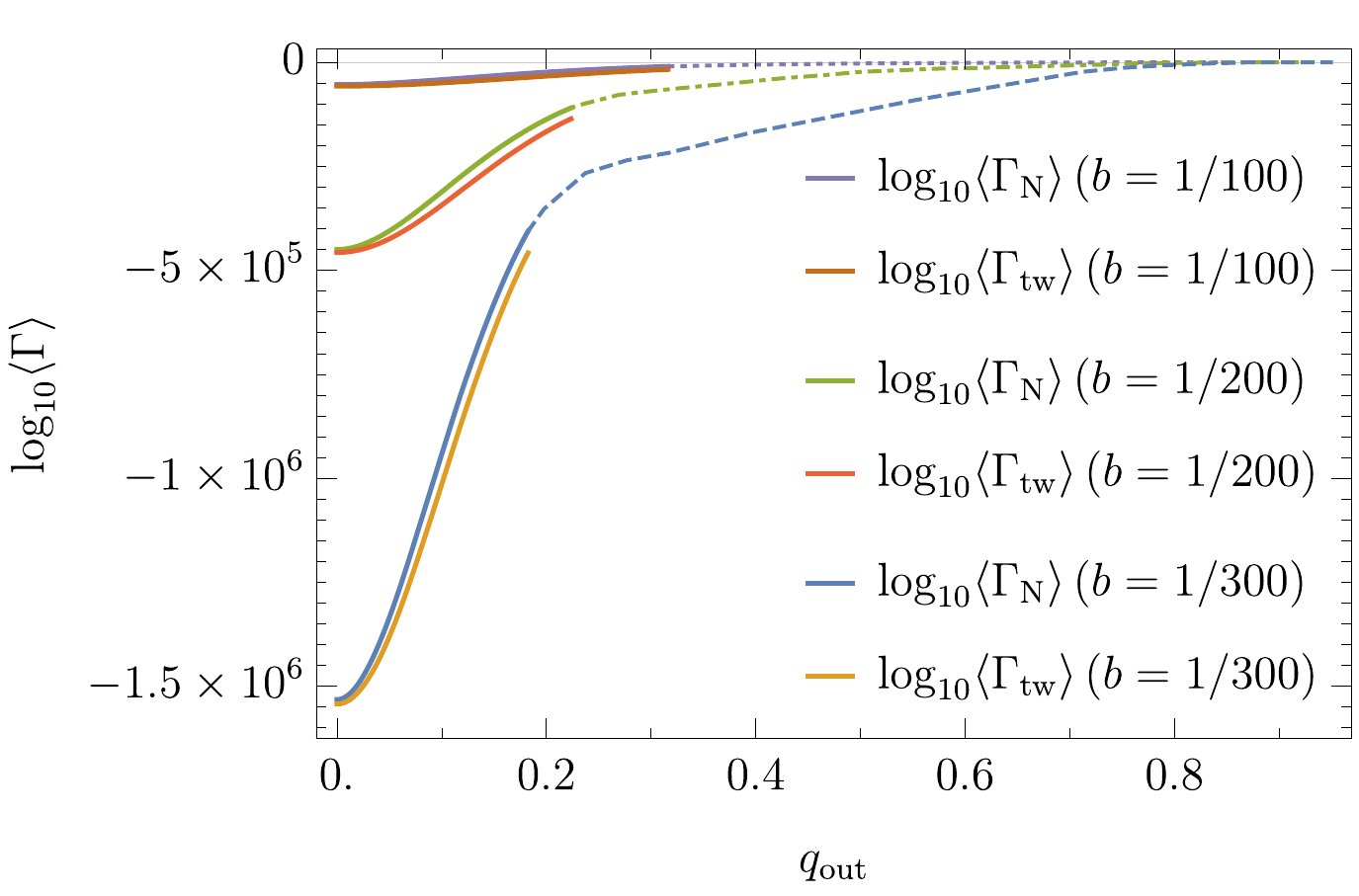}
\caption{
False vacuum averaged decay rate as a function of the amplitude of the oscillation of the background field $q_{\rm out}$. The dotted line show the region in which the bubble grows after nucleation according to our analytical results ($q_{\out} <\sqrt{2\alpha}$). The results with subscript $\rm N$ show numerically obtained action and dotted line takes into account bubbles with non-zero initial speed that grow according to our lattice evolution.
The values defining the potential were set to $g=1/10$ and $c=1$. 
\label{fig:decayrate}}
\end{figure}

Specifically, for oscillations with amplitude bigger than $q_{\rm out}>\sqrt{2\alpha_{\rm tw}}$ the smallest radius (and action) bubble corresponding to the extremum of the oscillation does not grow. However a slightly smaller bubble can grow due to non-zero initial background speed and these smaller bubbles drive the vacuum decay. More precisely, for oscillations larger than $\sqrt{2\alpha_{\rm tw}}$, we keep in our time average only the bubbles with boundary conditions~\eqref{eq:latticeboundary} which do grow. Such solutions are unlikely to be the result of the MPEP, since oscillations inside the bubble should be exponentially suppressed along this path as can be seen in~\eqref{eq:EoMQEucli}. However, they are still viable real time solutions of the classical equations of motion which furthermore satisfy energy conservation. The resulting average tunnelling rate provides an estimation of the tunnelling rate. Our analytical estimates cannot capture this effect as we have to assume small harmonic oscillations to obtain closed form results. However the two agree well as long as we are in the analytical result validity limit $q_{\rm out}<\sqrt{2\alpha_{\rm tw}}$, as shown in Figure~\ref{fig:decayrate}. 

\section{Summary and conclusions}\label{conclusions}
Tunnelling is one of the most profound phenomena in a quantum theory. It allows to transit from one local minimum to another one through a classically ``forbidden'' region. In quantum field theory it proceeds via the formation of a growing bubble of the new, lower vacuum from quantum fluctuations. This bubble subsequently grows in classical evolution. Its close cousin is bubble formation in first order phase transitions where a similar bubble is formed from the classical thermal fluctuations. Both situations
have in common that they start from a time-independent state corresponding to the minimum of the free energy.
Yet, it is easy to imagine a more complicated situation where the initial state itself exhibits significant time dependence. For example in cosmology classically evolving fields can be trapped in one of two (or several minima) by the damping effect of Hubble evolution. But even after the field cannot any more overcome the barrier, potentially large  residual oscillations remain. This leads us to the question how tunnelling proceeds from such a time-dependent initial state. In this note our aim was to take first steps to achieve such an understanding and provide estimates for the resulting tunnelling rates.

There are (at least) two sets of questions to be answered. 1) What is the bubble formation rate in such a state? What is its time dependence? 2) What is the time evolution after the initial bubble has formed? Does the bubble grow or can it actually collapse?

To be concrete in this note we looked at an initial situation where a scalar field initially oscillate in one of two available minima of a mexican hat potential with two non-degenerate minima. This situation has previously been studied by~\cite{KeskiVakkuri:1996gn}. However, our analysis differs by that we continue to imaginary time only in the inside region of the bubble. This leads to a significantly lower decay rate, but it also returns to the standard result in the limiting case of vanishing oscillations and large bubbles. A further crucial difference is that we study the subsequent bubble evolution. This turns out to be important even for the phase transition rate since there is a possibility of bubble collapse, i.e. the minimal action bubble may actually not introduce a phase transition.
\bigskip

Let us now turn to our actual calculations and survey the situation. 
We have approached the problem analytically as well as numerically.
To treat the problem analytically we have developed a version of the thin-wall approximation, taking into account the oscillating initial state and also allowing for a potentially oscillating solution inside the bubble. In the thin-wall limit the equation of motion is determined from a Lagrangian that includes a membrane term contracting the bubble, as well as pressure terms inside and outside the bubble. For small oscillations the main modifications arise from the pressure terms. As pressure is the {\emph{difference}} between kinetic and potential energy of the field the oscillating initial state corresponds to an oscillating pressure the effects of which will be averaged out over the field oscillations. The average pressure is then roughly equal to the vacuum-to-vacuum case. On the other hand, the initial radius of the bubble is fixed by the energy density which does depend on the amplitude of the oscillation. This distinction leads to the possible appearance of collapsing bubbles.

As long as the thin-wall limit is realised our numerical simulations agree very well with the thin-wall approximation. In particular as long as the thin-wall approximation is fulfilled the numerical results confirm the expectation that the field inside is close to the true vacuum. The agreement on the dividing line between growing and collapsing bubbles is very good for a sizeable range of parameters. Indeed for the growing case the bubbles themselves seem to be in rather good qualitative agreement with our analytical estimates. For contracting bubble we expect that our approximation of a constant field inside will break down and indeed we observe growing wiggles in the numerical simulation.

To summarise our results, as one may expect an oscillating initial field increases the tunnelling rate compared to a field sitting statically at the false vacuum. However, a new feature appears: bubbles with minimal action do not always grow -- there is the possibility that they can collapse. This has to be taken into account when the probability for a phase transition (and not just bubble nucleation rate) is to be calculated.

To get an appreciation of the challenges ahead let us look in a bit more detail at the different aspects contained within the general questions  posed above and delineate the open questions we still face.
To determine the time dependence we need to understand the following:
\begin{itemize}
\item{} The standard approach to tunnelling is based on a continuation to imaginary time. However, to do this smoothly in the usual simple manner requires a vanishing time derivative before and after the bubble has formed. However, for an oscillating initial state this is straightforward only at the turning point of the oscillation. A more general treatment is required and we will return to this in a forthcoming publication~\cite{2Dreduction}.
\item{} In the full quantum theory the initial state will have additional time dependence beyond the oscillations of the scalar field. Due to interactions the originally quantum fluctuations will grow. Effectively energy will be transferred from the coherent oscillations into fluctuations. Depending on the of values of the parameters of the model this may be a rather rapid process (cf. e.g.~\cite{Berges:2002cz}), leading to a significant time-evolution. This in turn can then lead to bubble formation based on these fluctuations. 
\end{itemize}
The bubble evolution will add further questions:
\begin{itemize}
\item{} If the minimal action bubbles contract, are there bubbles of possibly larger action that do grow instead. Those would then determine the phase transition rate.
\item{} Even if the bubbles collapse they may still be significant. They are a source of fluctuation growth, transferring energy from the coherent state to fluctuations. At the same time, and phenomenologically perhaps more important they are still a violent phenomenon, and therefore an interesting possible source of gravitational waves.
\end{itemize}
Due to the non-vanishing energy above the local ground state we also have to ask:
\begin{itemize}
\item{} Provided that the initial energy is big enough: can there also be ``tunnelling'' from an initial state close to the true vacuum to the false vacuum? In the thin-wall limit and assuming that oscillations are fast compared to all other changes of the bubble this seems difficult. As long as oscillations on both sides of the wall are harmonic, their contribution to the pressure differential average out and the average pressure difference is fixed by the energy difference between the true and the false vacuum. 
However, at present we have no stringent argument for the more general case.
\end{itemize}

To conclude, in many interesting situations, including in particular  in cosmology, phase transitions occurring via bubble formation may start from an initial state that is not in equilibrium and therefore explicitly time-dependent. Here we have taken first steps in understanding the bubble formation as well as its subsequent evolution. However, many open questions remain and provide opportunities for future investigations.

\bigskip
\bigskip
\noindent \textbf{Acknowledgments}
\medskip

\noindent
J.J.~would like to thank M.~G.~Schmidt for many useful discussions related to the subject of this paper. Moreover, he would like to thank J.~Berges, A.~Chatrchyan and K.~Boguslavski for always enjoyable discussions and collaboration on non-equilibrium physics.
The work of M.L. was supported by the ARC Centre of Excellence for Particle Physics at the Terascale (CoEPP) (CE110001104) and the Centre for the Subatomic Structure of Matter (CSSM).  M.L. was also supported in part by the Polish MNiSW grant IP2015 043174.
 L.D. is supported in part by the National Science Council (NCN) research grant No.~2015-18-A-ST2-00748. J.J. thankfully acknowledges support by the DFG TR33 ``The Dark Universe'' as well as the European Union Horizon 2020 research and innovation under the Marie Sklodowska-Curie grant agreement Numbers 674896 and 690575. 
\newpage
\appendix

\section*{Appendix}
\addcontentsline{toc}{section}{Appendices}
\section{Deriving the membrane action\label{app:mem}}

We start from the usual scalar field action (using the mostly minus signature for the metric),
\begin{align}
\label{eq:Sphi}
 S = \int dx^\mu \left[ \frac{1}{2} \d_\mu \phi \d^\mu \phi - V(\phi) \right] \ ,
\end{align}
where the $V(\phi)$ is a scalar potential featuring two quasi-degenerated vacua. Our goal is to use our assumption on the spatial structure of the classical solitonic solution to express this action as an integral over time with the radius of the bubble as the only free parameter. First, we assume a spherically symmetric field configurations, so that~\eqref{eq:Sphi} can be written as 
\begin{align}
\label{eq:Sphi2}
 S = 4 \pi \iint dt d\rho \rho^2 \left[ \frac{1}{2} (\dot{\phi}^2 - (\d_\rho \phi)^2) - V(\phi) \right] \ ,
\end{align}
where we have used $\rho^2 = x^2+y^2+z^2$, and denoted the time derivative with a dot. We will suppose the presence of a ``wall'' around the radial position $R(t)$ and an homogeneous region where $ \d_\rho\phi \simeq 0$ in the rest of the space. Noting the field within (outside) the wall as $ c+ \phiin$ ($-c+\phiout$) we can decompose~\eqref{eq:Sphi2} as
\begin{align}
\label{eq:Sphitmp}
 S = 4 \pi \left[ \iint_{wall} \! \! dt d\rho~ \rho^2 \left(\frac{1}{2} (\dot{\phi}^2 - {\d_\rho \phi}^2) - V(\phi) \right) + \int dt \frac{R^3}{3} p \right] \ ,
\end{align}
where we have defined the differential pressure as
\begin{align}
  p = p_{in} -p_{\out} \equiv \left(\frac{1}{2} \dot{\varphi}_{in}^2 - V(c+ \phiin) \right) - \left(\frac{1}{2} \dot{\varphi}_{\out}^2 - V(-c+\phiout) \right) \ .
\end{align}
Here we have not included the terms from the infinite spatial volume outside the bubble, as in our approximation they do not affect the evolution of the interiors as well as the bubble position.

Let us now concentrate on the wall region. In Minkowski space, we can write the unit vectors perpendicular and parallel to the wall as function of the time and radial vierbein elements $\eb_t$ and $\eb_\rho$:
\begin{align*}
\eb_\perp &=\frac{1}{\sqrt{1-\dot{R}^2}} (-\dot{R} \eb_t + \eb_\rho) &&& \eb_\parallel &= \frac{1}{\sqrt{1-\dot{R}^2}} (\eb_t + \dot{R}  \eb_\rho)\ .
\end{align*}
We will then make the assumption that in thin-wall region
\begin{align*}
 \d_\perp \phi ~\equiv~ \eb_\perp^\mu d_\mu \phi ~\gg~  \d_\parallel \phi ~\equiv~ \eb_\parallel^\mu d_\mu \phi \ .
\end{align*}
In particular, the equations of motion in this region can be written as
\begin{align*}
-|\eb_\perp|^2 \d^2_\perp \phi ~=~ \d^2_\perp \phi ~=~ V'(\phi) \ ,
\end{align*}
where we have neglected the first order ``friction'' terms\footnote{which takes in general the form $\frac{\ddot{R} R + 2(1-\dot{R}^2)}{(1-\dot{R}^2)^{3/2}} \d_\perp \phi$} in front of the second order one. This equations is easily integrated from the inside of the wall to a point within the wall into
\begin{align}
 (\d_\perp \phi)^2 - \left. (\d_\perp \varphi)^2 \right|_{\in} = 2 V(\phi) - 2 V(c+\phiin) \ .
\end{align}
We then neglect the derivative contributions from inside the wall in this equation as part of the ``thin wall'' limit: the field variation is concentrated within the wall. Further assuming that $V(c+\phiin) $ is close to the second minimum where the potential energy should be negligible compared to the potential energy in the wall we obtain:
\begin{align}
 (\d_\perp \phi)^2  = 2 V(\phi)  \ .
\end{align}

We are now ready to calculate the tension term in~\eqref{eq:Sphitmp}:
\begin{align*}
  \iint_{wall}\! \! dt d\rho~ \rho^2  \left(\frac{1}{2} (\dot{\phi}^2 - {\d_\rho \phi}^2) - V(\phi) \right) &= \int dt \int_{wall} \! \! d\rho~ \rho^2 [-\frac{1}{2} (\d_\perp \phi)^2 - V(\phi)] \\
  &= \int dt R(t)^2 \int_{wall} \! \! d\rho~  [-(\d_\perp \phi)^2] \\
   &= -\int dt~ R(t)^2  \int_{-c+\phiout}^{c+\phiin} d\phi \frac{\sqrt{1-\dot{R}^2}}{\d_\perp \phi} (\d_\perp \phi)^2   \\
 &= -\int dt~ R(t)^2 \sqrt{1-\dot{R}^2}  \int_{-c+\phiout}^{c+\phiin} d\phi \sqrt{2 V(\phi)} \\
  &\equiv -\int dt~ R^2 \sqrt{1-\dot{R}^2}\ \sigma 
   \end{align*}
where we have defined the tension as
\begin{align}
\sigma =  \int_{-c+\phiout}^{c+\phiin} d\phi \sqrt{2 V(\phi)}  \ .
\end{align}

\newpage

\bibliographystyle{utphys}
\bibliography{tunneling}

\end{document}